\documentclass{elsart}
\usepackage{graphicx}
\begin{document}
\begin{frontmatter}
\title{Correlations beyond the mean field in Magnesium isotopes:
Angular momentum projection and configuration mixing.}
\author{R.Rodr\'{\i}guez-Guzm\'an\thanksref{PA1}},
\author{J. L. Egido} and
\author{L. M. Robledo\corauthref{cor1}}
\address{Departamento de F\'\i sica Te\'orica C--XI, Universidad
Aut\'onoma de Madrid, E--28049 Madrid, Spain}
\corauth[cor1]{Corresponding author. Email address: \texttt{Luis.Robledo@uam.es}}
\thanks[PA1]{Present address: Institut f\"ur Theoretische Physik der
Universit\"at T\"ubingen, 
Auf der Morgenstelle 14, D-72076 T\"ubingen, Germany.}

\begin{abstract}
%
The quadrupole deformation properties of the ground and low-lying excited states
of the even-even Magnesium isotopes with N ranging from 8 to 28 have been 
studied in the framework of the angular
momentum projected generator coordinate method with the Gogny force.  It is
shown that the N=8 neutron magic number is preserved (in a dynamical sense) in
\nuc{20}{Mg} leading to a spherical ground state. For the magic numbers N=20
and N=28 this is not the case and prolate deformed ground states are obtained. 
The method yields values of the two neutron separation energies which are in much
better agreement with experiment than those obtained at the mean field level.
It is also obtained that \nuc{40}{Mg} is at the neutron dripline. Concerning
the results for the excitation energies of the $2^+$ excited states and their
transition probabilities to the ground state we observe a good agreement with
the available experimental data. On the theoretical side, we also present a
detailed justification of the prescription used for the density dependent part
of the interaction in our beyond-mean-field calculations. 
\end{abstract}
\end{frontmatter}

\section{Introduction}
%

Thanks to the new breed of Radioactive Isotope Beam (RIB) facilities set up in the
last years all over the world the study of exotic systems far away from the
valley of stability has become one of the main topics in today's nuclear
physics.  On the neutron  rich side, because neutrons do not carry electric
charge,  the dripline can extend far away from the valley of stability. Many
neutron rich regions have been the  subject of detailed studies in recent years
and just to give an example one should mention the regions around the magic
neutron numbers N=20 and N=28. These exotic  systems show many new interesting
features which deserve a careful study 
\cite{ref_a,ref_b,ref_c,ref_d,ref_e,ref_f,ref_g,ref_h,ref_i,ref_j}. Among them,
the weakening of the neutron magic numbers N=20 and N=28 in some nuclei is
probably the most challenging to study.  

It was in the neutron rich region N $\approx$ 20 where the breaking of a shell
closure was first detected \cite{ref_k} and  related with a shape transition to
prolate deformed shapes  \cite{ref_l}. Ground  state deformations in this
region easily explain the anomalous isotope shift observed in \nuc{31}{Na}
\cite{ref_m} as well as the  decrease in  the two neutron separation energy
$S_{2N}$ in \nuc{31,33,35}{Na} and \nuc{30}{Ne} \cite{ref_k,ref_m}.  A large
deformation has also been inferred in the N=20 nucleus \nuc{32}{Mg} 
through the measurements of $B(E2,0_{1}^{+} \rightarrow
2_{1}^{+})$ transition probabilities \cite{ref_o}. In addition, the low excitation energy 
of the first $2_{1}^{+}$  excited state \cite{ref_p,ref_q} in \nuc{32}{Mg} and
the ratio $E(4^+)/E(2^+)=2.6$ are consistent with the expectations for a
rotational state. Recently, evidence for a rotational band in \nuc{31}{Na} has
also been obtained \cite{Prity.00}. Although indirect, another  evidence for the
deformed nature of the \nuc{32}{Mg} ground state comes from the strong ground
state deformation of \nuc{34}{Mg} inferred from the large value of the
$B(E2,0_{1}^{+} \rightarrow 2_{1}^{+})$ transition probability  measured
\cite{Iwa.01} in this nucleus and the $E(4^+)/E(2^+)$ ratio of 3.18 
\cite{ref_u,ref_v} which is very close to the rotational limit.
Further evidence for deformed ground states in
the neutron rich Mg isotopes have also been obtained experimentally
\cite{ref_s,ref_t,BE2_26_28Mg}. In addition, the  quadrupole collectivity  in
\nuc{32-38}{Si} was experimentally studied in \cite{Ibbo98}. Finally, an
extensive account of older experimental results can be found in Refs.
\cite{Endt90,Endt98}.

In the framework  of the mean field approximation the \nuc{32}{Mg} ground
state  has been found to be spherical (see for example
\cite{ref_aa,ref_bb,ref_cc,ref_dd,ref_w,ref_x}) but it has also  been argued
that dynamical correlations may play an important role in this and other nuclei
of the region. In particular, the results presented in 
\cite{ref_l,ref_y,ref_z,ref_ee,ref_ff,ref_gg,ref_hh,ref_ii} gave initial hints
concerning  the importance of the restoration of symmetries and configuration
mixing in some of the exotic  nuclei considered. From the Shell Model point of 
view, the large deformations around N=20 have been explained by invoking
excitations  of pairs of neutrons across the N=20 shell gap (2 $\hbar \omega$
configurations). In a  small set of nuclei (the  {\it island of inversion}) the
coexisting 2 $\hbar \omega$  intruder states actually fall below the normal 0
$\hbar \omega$ states  and become the ground state configurations. In this
way  Shell Model calculations with different levels of complexity 
\cite{ref_jj,ref_kk,ref_ll,ref_mm,ref_nn,ref_oo,ref_pp} have been able to
explain the increased quadrupole collectivity around N=20.

The existence of anomalies in shell closures, as already found in nuclei with N
$\approx$ 20, is one of the main features for the experimental and  theoretical
studies carried out up to now in the neutron rich nuclei with  N$ \approx$ 28.
The  $\beta$-unstable nuclei around \nuc{44}{S} have  attracted particular
interest since these nuclei play an important role in the  nucleosynthesis
process. The experimental study of the   $\beta$-decay properties of 
\nuc{44}{S}  (N=28) indicated \cite{ref_qq,ref_rr} that this nucleus is
deformed. This has been confirmed in subsequent intermediate energy Coulomb
excitation studies \cite{ref_ss,ref_tt,ref_uu}. More recent experiments  on
nuclei of this region \cite{ref_yy,ref_zz} suggest that \nuc{44}{S} is indeed 
a deformed nucleus but with strong  shape coexistence making more challenging
its theoretical description. 

The region N $\approx$ 28  has been studied using the  Skyrme Hartree-Fock (HF)
model  and the relativistic mean field approximation    (RMF)
\cite{ref_vv,ref_ww,ref_xx,ref_LALA_ROBLE,ref_aaa}. These studies also showed
the onset of deformation for some nuclei with N $\approx$ 28. Surprisingly, in 
none of these mean field calculations the rotational energy correction has been
considered in spite that it is known to be  a key ingredient for a proper
description of the energy landscape as a function of the  quadrupole
deformation. However, in the  mean field calculations of \cite{ref_z} with
several parameterizations of the Skyrme  interaction, in the calculation
reported in \cite{ref_hh} in the framework of the  Bohr Hamiltonian method with
the Gogny interaction or in our  recent study  \cite{N=28} of the phenomenology
of quadrupole collectivity around N=28 the rotational energy correction was
explicitly taken into account either in an approximate way or exactly. The
conclusions extracted from those calculations concerning the erosion of the 
shell closure at N=28 do not differ  qualitatively (but they do in the
quantitative side) from the ones extracted from the calculations without the
rotational energy correction. This indicates that the N=28 shell closure can be
more easily broken than the N=20 one (see also \cite{Cottle-Kemper}). From the
Shell Model point of view  an erosion of the N=28 shell closure  in the sulphur
isotopes has also been found in calculations with the full  $sd$ shell for
protons and the full $pf$ shell for neutrons (see, for example
\cite{ref_bbb}).  

The purpose of this paper is to perform  a systematic study, in the nuclei
\nuc{20-40}{Mg},  of  the  quadrupole deformation properties of the ground and
excited states. This isotopic chain contains the three spherical  magic numbers
N=8, N=20 and N=28 and therefore it is a good testing ground  to examine
both the systematic of deformation and the possible erosion of  spherical shell
closures. Besides, most of  the considered nuclei  are  examples where the
mean field energy landscape as a function of quadrupole  deformation shows 
prolate and oblate minima which are  practically degenerate in energy.
Therefore, the correlation energies  associated with the restoration of broken
symmetries  (mainly the rotational symmetry) and the  collective quadrupole
motion have to be considered at the same time. In the present  study both effects are taken into
account  in the framework of the  Angular Momentum Projected Generator
Coordinate Method \cite{ref_gg,ref_ii,N=28,ref_32S}. The main reason to perform
an exact angular momentum projection is that the usual approximations, like the
Gaussian Overlap Approximation (GOA), used to compute the  rotational
energy correction as well as the $B(E2)$ transition  probabilities are not
expected to work well \cite{ref_ff,ANGEL_24Mg} at least for  some of the 
nuclei considered in this study.

We have used in our calculations the Gogny  interaction \cite{D1} with the
parameterization $D1S$ \cite{D1S_1}. The use of the Gogny interaction in this
systematic  study is justified not  only by the fact that this interaction
provides reasonable results  for many nuclear properties all over the nuclear
chart, but also by the  good description of the phenomenology of quadrupole
collectivity  in the regions N $\approx$ 20 and N $\approx$ 28 obtained
recently in the  same framework as the one used in this study
\cite{ref_ff,ref_gg,ref_ii,N=28} as well as in the context of the Bohr
collective Hamiltonian  \cite{ref_y,ref_hh}. As the results presented  in this
paper will show, this force also provides reasonable  results for the
systematic  description of the phenomenology  of  quadrupole collectivity  in
the chain of Magnesium isotopes. 

The paper is organized as follows: in Section 2, we describe the theoretical 
formalisms used in the present paper. These formalisms are the  Angular
Momentum Projection (AMP) and the  Angular Momentum Projected Generator
Coordinate Method (AMPGCM) \cite{ref_ff,ref_gg,ref_ii,N=28,ref_32S}  with the
axially symmetric quadrupole moment as generating coordinate. The angular
momentum projection technique is presented in section 2.1 where also special
attention is paid to the specific problems appearing in the case of a density
dependent interaction like the Gogny force. In section 2.2 we present the
simplified expressions corresponding to the results of 2.1 in the case of
axially symmetric intrinsic wave functions. Configuration mixing of angular
momentum projected  Hartree-Fock-Bogoliubov states   is presented in section
2.3.  There we also present  the expressions to compute both  transition
probabilities and spectroscopic quadrupole moments (for more details the reader
is referred to Appendix A). The results of our calculations for several
Magnesium isotopes are discussed  in Section 3. In section 3.1 we present the
results for the underlying mean field studies. In section  3.2 the topological
changes introduced in the mean field potential energy surfaces due to the exact
restoration of the rotational symmetry are presented and then  the results of
the angular momentum projected configuration mixing calculations are discussed
and compared with the available experimental data and other theoretical
calculations. Finally Section 4 is devoted to some concluding remarks.

\section{Theoretical framework: Angular momentum projection and configuration 
mixing}
%

As the results of the next sections will show, the mean field approximation is
just an  starting point and additional correlations have to be incorporated in
order to properly  describe the considered nuclei. The small energy differences
observed between the  coexisting minima indicate that many body effects beyond
the mean field, like the  restoration of the rotational symmetry and/or
quadrupole fluctuations, may change the  conclusions extracted from the raw
Hartree-Fock-Bogoliubov approximation. The restoration  of the rotational
symmetry leads to an energy gain (the so called rotational  energy correction)
which is proportional to the quadrupole deformation of the intrinsic state and
ranges  from zero (spherical intrinsic state) to a few MeV for typical well
deformed configurations  in the region of the nuclear chart considered in this
study. Therefore,  the consideration of this effect might play a key role for a
more qualitative and quantitative  description of the nuclei we are interested
here. In addition, it is very important   to consider the effect of quadrupole
mixing  \cite{ref_ee,ref_ff,ref_gg,ref_ii,N=28,ANGEL_24Mg}) in those cases
where the  angular momentum projected potential energy surfaces show important
topological changes  compared with the corresponding mean field surfaces. 

The theoretical background for the restoration of rotational symmetry  is
treated extensively in the literature (see, for example, Refs.
\cite{ref_RING,KARL_1,KARL_2,HARA_RING,ref_HARA_SUN,ref_ENAMI}) and also very
good books and papers concerning the treatment of  correlations beyond mean
field in the framework of the Generator Coordinate Method (GCM) are 
available (see, for example,  Refs. \cite{ANGEL_24Mg,ref_RING,GCM}).
Thus we are not going  to dwell on these details  here. In spite of this we
believe that it is useful to present  a short outlook of  our procedure to
carry out both angular momentum and configuration mixing  in the case of a
density  dependent interaction like the Gogny force. In addition, the issue of
which density dependence has to be used in these calculations beyond mean field
will be discussed in detail and strong arguments in favor of the prescription
used will be given.

\subsection{Angular momentum projection with density 
dependent interactions.}
  
Let us consider a set of mean field states $\mid \varphi(\vec{q}) \rangle$ depending 
on the  parameters 
$\vec{q}=\{q_{20},q_{22},q_{40},\dots\}$ 
which define the corresponding mean field
configurations. From this set we can build another set of states where
the rotational symmetry is restored
\cite{ref_RING,KARL_1,KARL_2,HARA_RING,ref_HARA_SUN,ref_ENAMI}

\begin{eqnarray} \label{PROJWF}
\mid \Psi_{IM}(\vec{q})\rangle &=&
\sum_{K} g_{K}^{I}(\vec{q}) \hat{P}_{MK}^{I} 
\mid \varphi(\vec{q})\rangle.
\end{eqnarray}
The operator $\hat{P}_{MK}^{I}$ is the angular momentum
projection operator \cite{Pier_1} given by

\begin{eqnarray} \label{AMProj}
\hat{P}_{M K}^{I} &=&
\frac{2I+1}{8{\pi}^{2}}
\int d\Omega 
\mathcal{D}_{M K}^{I *} (\Omega)
\hat{R}(\Omega)
\end{eqnarray}

with $\Omega$ representing the set of the three Euler angles 
$\left(\alpha, \beta, \gamma  \right)$, $\mathcal{D}_{M K}^{I} (\Omega)$ is
the well known Wigner function \cite{Varsh.88} and 
$\hat{R}(\Omega)= 
e^{-i \alpha \hat{J}_{z}} e^{-i \beta \hat{J}_{y}} e^{-i \gamma \hat{J}_{z}}$
is the rotation operator. The energy $E^{I}(\vec{q})$  is given by

\begin{equation} \label{PROJENER} 
E^{I}(\vec{q}) = \frac{
\langle \Psi_{IM}(\vec{q}) \mid \hat{H} \mid \Psi_{IM}(\vec{q}) \rangle}{
\langle \Psi_{IM}(\vec{q}) \mid \Psi_{IM}(\vec{q}) \rangle}
\end{equation}

The above procedure is based on the assumption that the hamiltonian
is rotational invariant and therefore the intrinsic energy \( \langle \varphi (\vec{q})|\hat{H}|\varphi (\vec{q})\rangle  \)
is independent of the orientation of the intrinsic wave function \( |\varphi (\vec{q})\rangle  \),
that is 
\begin{equation}
\label{rotinv}
\langle \varphi (\vec{q})|\hat{H}|\varphi (\vec{q})\rangle =\langle \varphi (\vec{q})|\hat{R}^{\dagger }(\Omega )\hat{H}\hat{R}(\Omega )|\varphi (\vec{q})\rangle 
\end{equation}

However, when dealing with density dependent (DD) forces the above
assumption apparently breaks down as, in general, the density dependent
term is not rotational invariant. This apparent paradox can be solved
if density dependent hamiltonians are thought not as  genuine hamiltonians
but rather as devices to get an elaborated energy functional of the
density. With this point of view the right hand side of Eq. (\ref{rotinv})
has to be treated as
\begin{equation}
\label{rotinv1}
\langle \varphi (\vec{q})|\hat{R}^{\dagger }(\Omega )\hat{H}
[\bar{\rho}_{\Omega,\Omega }(\vec{r})]\hat{R}(\Omega )|\varphi (\vec{q})\rangle 
\end{equation}
 where the interaction depends on the rotated density 
\begin{equation}
\label{rotdens}
\bar{\rho}_{\Omega,\Omega }(\vec{r})=\langle \varphi (\vec{q})|\hat{R}^{\dagger }(\Omega )\hat{\rho }(\vec{r})\hat{R}(\Omega )|\varphi (\vec{q})\rangle 
\end{equation}
 with the density operator defined in the usual way
 \begin{equation}
\label{densop}
\hat{\rho }(\vec{r})=\sum ^{A}_{i=1}\delta (\vec{r}-\vec{r}_{i})
\end{equation}
 Before proving that Eq.  (\ref{rotinv}) holds for density dependent
interactions let us present some basic results. First, we introduce
the $ 3\times 3 $ unitary rotation matrix $ \mathcal{R} ( \Omega ) 
\equiv \mathcal{R}_z (\alpha) \mathcal{R}_y (\beta)\mathcal{R}_z (\gamma)$ that
appears when eigenstates $\mid \vec{r} \rangle$ of the position operator are
rotated
\begin{equation}\label{Rotr}
\hat{R} (\Omega) \mid \vec{r} \rangle = \mid \mathcal{R}\vec{r} \rangle. 
\end{equation}
This is the same matrix that appears when  vector operators 
(like the position operator) are rotated
\begin{equation} \label{Rotrop}
\hat{R}^{\dagger }(\Omega )\hat{\vec{r}}\hat{R}(\Omega )=\mathcal{R}(\Omega )
\hat{\vec{r}}
\end{equation}
Now we have to consider the role played by the density operator. Its 
mean value (or overlap)
appears in the density dependent part of the hamiltonian and therefore the
``parameter" $\vec{r}$ of Eq. (\ref{densop}) has to be treated as the position
operators in that case. As a consequence the action of the rotation operator on
$\vec{r}$ has to be considered
 \begin{equation}
\label{rotdens1}
\hat{R}^{\dagger }(\Omega )\hat{\rho }(\vec{r})\hat{R}(\Omega )=
\hat{\rho }(\mathcal{R}(\Omega )\vec{r})=
\sum ^{A}_{i=1}\delta (\mathcal{R}(\Omega )\vec{r}-\vec{r}_{i})=
\sum ^{A}_{i=1}\delta (\vec{r}-\mathcal{R}^{\dagger }(\Omega )\vec{r}_{i})
\end{equation}
as well as its action on the internal coordinates \( \vec{r}_{i} \)
\begin{equation}
\label{rotdens2}
\hat{R}^{\dagger }(\Omega )\hat{\rho }(\vec{r})\hat{R}(\Omega )=
\sum ^{A}_{i=1}\delta (\vec{r}-\mathcal{R}(\Omega )\vec{r}_{i})=
\sum ^{A}_{i=1}\delta (\mathcal{R}^{\dagger }(\Omega )\vec{r}-\vec{r}_{i})=
\hat{\rho }(\mathcal{R}^{\dagger }(\Omega )\vec{r}).
\end{equation}
With these results we have for the rotated density of Eq. (\ref{rotdens}) that 
$$
\hat{R}^{\dagger }(\Omega )\bar{\rho}_{\Omega,\Omega }(\vec{r})\hat{R}(\Omega )=
\langle \varphi (\vec{q})|\hat{R}^{\dagger }(\Omega )
\hat{\rho }(\mathcal{R}(\Omega )\vec{r})\hat{R}(\Omega )
|\varphi (\vec{q})\rangle =
\langle \varphi (\vec{q})|\hat{\rho }(\vec{r})|\varphi (\vec{q})\rangle =
\rho(\vec{r})
$$
 The first identity is the result of applying Eq. (\ref{rotdens1})
whereas the second is the result of applying Eq. (\ref{rotdens2}).
The result thus obtained shows that even for density dependent forces
Eq. (\ref{rotinv}) holds.

In order to evaluate the projected energy we need to know the Hamiltonian
overlaps \( \langle \varphi \mid \hat{R}^{\dagger }(\Omega )\hat{H}\hat{R}(\Omega ')\mid \varphi \rangle  \).
For density independent and rotational invariant interactions the
hamiltonian overlap is simply given by \( \langle \varphi \mid \hat{H}\hat{R}(\Omega '-\Omega )\mid \varphi \rangle  \).
For density dependent interactions this quantity is not well defined
and therefore we have to look for a prescription for the density dependent
term. The most popular prescription so far is to use the density
\begin{equation}
\overline{\rho }_{\Omega ,\Omega '}(\vec{r})=
\frac
{\langle \varphi (\vec{q})\mid \hat{R}^{\dagger }(\Omega )\hat{\rho }(\vec{r})\hat{R}(\Omega ')\mid \varphi (\vec{q})\rangle }
{\langle \varphi (\vec{q})\mid \hat{R}^{\dagger }(\Omega )\hat{R}(\Omega ')\mid \varphi (\vec{q})\rangle }
\label{HAMDENJ} 
\end{equation}
 in the density dependent part of the interaction. Several arguments
to show the credibility of this prescription have been given. For
example, for those Skyrme forces, with a linear dependence in the density,
that can be cast as a three body force the hamiltonian overlap can
be computed without ambiguity and the resulting density dependent
term depends on the density of Eq. (\ref{HAMDENJ}). Arguments
based on the extended Wick theorem (the fundamental tool to compute the overlap of operators
between different product wave functions) are also very popular. Apart from
the previous arguments, it is mandatory for the prescription to
be used that it fulfills the following two requirements: First, it should
not carry angular momentum or, in other words, the following identity
should be fulfilled
\begin{equation}
\label{Require}
\langle \Psi_{I_1M_1}(\vec{q})\mid \hat{H}
\mid \Psi_{I_{2}M_{2}}(\vec{q})\rangle =
\delta _{I_{1}I_{2}}\delta _{M_{1}M_{2}}
\langle \Psi_{I_{1}M_{1}}(\vec{q})\mid \hat{H}
\mid \Psi_{I_{1}M_{1}}(\vec{q})\rangle .
\end{equation}
 Second, the prescription must lead to a real projected energy (notice
that, in general, the density of Eq. (\ref{HAMDENJ}) is a complex
quantity).

The first requirement (Eq. (\ref{Require})) is a direct consequence
of the property 
\begin{eqnarray}
\overline{\rho }_{\Omega ,\Omega '}(\vec{r}) & = & 
\frac{\langle \varphi (\vec{q})\mid \hat{R}^{\dagger }(\Omega )\hat{\rho
}(\vec{r})\hat{R}(\Omega ')\mid \varphi (\vec{q})\rangle }{\langle \varphi
(\vec{q})\mid \hat{R}^{\dagger }(\Omega )\hat{R}(\Omega ')\mid \varphi
(\vec{q})\rangle } \\ \nonumber
& = & \frac{\langle \varphi (\vec{q})\mid \hat{\rho }(\mathcal{R}^{\dagger }(\Omega
)\vec{r})\hat{R}(\Omega '-\Omega )\mid \varphi (\vec{q})\rangle }{\langle
\varphi (\vec{q})\mid \hat{R}(\Omega '-\Omega )\mid \varphi (\vec{q})\rangle } \\ \nonumber
& = & \overline{\rho }_{0,\Omega '-\Omega }(\mathcal{R}^{\dagger }(\Omega )\vec{r})
\end{eqnarray}
which is easily obtained using Eq. (\ref{rotdens2}). Using this result
we obtain

\begin{equation}
\hat{R}^{\dagger }(\Omega )\overline{\rho }_{\Omega ,\Omega '}(\vec{r})\hat{R}(\Omega )=\overline{\rho }_{0,\Omega '-\Omega }(\vec{r}).
\end{equation}
This property allows to write the numerator of the projected energy as

\begin{eqnarray}
\int d\Omega d\Omega '\mathcal{D}_{M_{1}K_{1}}^{I_{1} }(\Omega  )
                      \mathcal{D}_{M_{2}K_{2}}^{I_{2}*}(\Omega ')
		      \langle \varphi (\vec{q})\mid \hat{R}^{\dagger }(\Omega )
		      \hat{H}\left[ \overline{\rho }_{\Omega ,\Omega '}(\vec{r})\right] 
		      \hat{R}(\Omega ')\mid \varphi (\vec{q})\rangle = 
		      & \nonumber  \\
\int d\Omega d\Omega '\mathcal{D}_{M_{1}K_{1}}^{I_{1} }(\Omega )
                      \mathcal{D}_{M_{2}K_{2}}^{I_{2}*}(\Omega ')
		      \langle \varphi (\vec{q})\mid \hat{H}
		      \left[ \overline{\rho }_{0,\Omega '-\Omega }(\vec{r})\right] 
		      \hat{R}(\Omega '-\Omega )\mid \varphi (\vec{q})\rangle = 
		      & \nonumber \\
\sum _{m}\int d\xi    \mathcal{D}_{M_{1}K_{1}}^{I_{1}}(\xi )
                      \mathcal{D}_{M_{2}m}^{I_{2}*}(\xi )
		      \int d\Omega 
		      \mathcal{D}_{mK_{2}}^{I_{2}*}(\Omega )\langle \varphi (\vec{q})\mid 
		      \hat{H}\left[ \overline{\rho }_{\Omega }(\vec{r})\right] 
		      \hat{R}(\Omega )\mid \varphi (\vec{q})\rangle = 
		      & \nonumber \\
\frac{8{\pi }^{2}}{2I_{1}+1}{\delta }_{M_{1},M_{2}}{\delta }_{I_{1},I_{2}}
\int d\Omega \mathcal{D}_{K_{1}K_{2}}^{I_{1}*}(\Omega )\langle 
\varphi (\vec{q})\mid \hat{H}\left[ \overline{\rho }_{\Omega }(\vec{r})\right] 
\hat{R}(\Omega )\mid \varphi (\vec{q})\rangle  
& \nonumber  
\end{eqnarray}
 with 
 $$ 
 \overline{\rho }_{\Omega }(\vec{r})\equiv \overline{\rho }_{0,\Omega }(\vec{r}) 
 $$
given by Eq. (\ref{HAMDENJ}). 

The projected energy Eq. (\ref{PROJENER}) can now be written as
\begin{equation}\label{PROJENER1}
E^{I}(\vec{q})=
\frac
{\sum _{KK^{'}}g_{K}^{I*}(\vec{q})g_{K^{'}}^{I}(\vec{q})
\int d\Omega \mathcal{D}_{KK^{'}}^{I *}(\Omega )h(\Omega )}
{\sum _{KK^{'}}g_{K}^{I*}(\vec{q})g_{K^{'}}^{I}(\vec{q})
\int d\Omega \mathcal{D}_{KK^{'}}^{I *}(\Omega )n(\Omega )}
\end{equation}
 with
\begin{equation}
h(\Omega )=\langle \varphi (\vec{q})\mid \hat{H}\left[ \overline{\rho }_{\Omega }(\vec{r})\right] \hat{R}(\Omega )\mid \varphi (\vec{q})\rangle 
\end{equation}
and
\begin{equation}
n(\Omega )=\langle \varphi (\vec{q})\mid \hat{R}(\Omega )\mid \varphi (\vec{q})\rangle .
\end{equation}
 This result shows that the energy is a quantity independent of the
quantum number \( M \) and therefore of the orientation of the laboratory
reference frame. 

In order to ensure the reality of the projected energy 
\cite{KARL_1,KARL_2,HARA_RING} we have to show that the  property 
\footnote{The same property holds for the norm overlap
$n(\Omega)$ and also for the neutron and proton overlap
$N(\Omega) = 
\langle \varphi(\vec{q}) \mid    
\hat{N} \hat{R}(\Omega) \mid \varphi(\vec{q}) \rangle$ 
 and $Z(\Omega) = 
\langle \varphi(\vec{q}) \mid    
\hat{Z} \hat{R}(\Omega) \mid \varphi(\vec{q}) \rangle$. Here
 and in the following we have  mainly concentrated 
on the hamiltonian overlap
$h(\Omega)$.
}

\begin{equation} \label{hstar}
h^{*}(\Omega) = h(-\Omega)
\end{equation}
holds even for density dependent forces. Here we would like to remark again 
that the mixed density of Eq. (\ref{HAMDENJ}) is not in general  a real
quantity and is not necessarily  positive definite.
The complex conjugate
density in the density dependent term is given by

\begin{eqnarray}
\overline{{\rho}}^{*}_{\Omega}(\vec{r}) &=&
\frac{
\langle \varphi(\vec{q}) \mid \hat{R}^{\dagger}(\Omega) \hat{\rho}(\vec{r}) 
\mid \varphi(\vec{q}) \rangle}
{\langle \varphi(\vec{q}) \mid   \hat{R}^{\dagger}(\Omega)
\mid \varphi(\vec{q}) \rangle} \nonumber\\
& = & 
\frac{
\langle \varphi(\vec{q}) \mid  \hat{\rho}(\mathcal{R}^\dagger(\Omega)\vec{r}) \hat{R}(-\Omega)
\mid \varphi(\vec{q}) \rangle}
{\langle \varphi(\vec{q}) \mid   \hat{R}(-\Omega)
\mid \varphi(\vec{q}) \rangle}
\nonumber\\
&=&
{\overline{\rho}}_{-\Omega}(\mathcal{R}^\dagger (\Omega)\vec{r}).
\nonumber\\
\end{eqnarray}
Using this result we have

\begin{eqnarray}
h^{*}(\Omega) &=& 
\langle \varphi(\vec{q}) \mid \hat{R}^{\dagger}(\Omega) \hat{H} 
\left[\overline{{\rho}}^{*}_{\Omega}(\vec{r})\right]
\mid \varphi(\vec{q}) \rangle
\nonumber\\
&=& 
\langle \varphi(\vec{q}) \mid \hat{R}^{\dagger}(\Omega) \hat{H} 
\left[\overline{{\rho}}^{*}_{\Omega}(\vec{r})\right]
\hat{R}(\Omega) \hat{R}(- \Omega) 
\mid \varphi(\vec{q}) \rangle
\nonumber\\
&=& 
\langle \varphi(\vec{q}) \mid \hat{R}^{\dagger}(\Omega) \hat{H} 
\left[{\overline{\rho}}_{-\Omega}(\mathcal{R}^\dagger (\Omega)\vec{r})\right]
\hat{R}(\Omega) \hat{R}(- \Omega) 
\mid \varphi(\vec{q}) \rangle
\nonumber\\
\end{eqnarray}

Now, using
\begin{equation}
\hat{R}^{\dagger}(\Omega) 
\overline{{\rho}}_{-\Omega}(\mathcal{R}^\dagger (\Omega) \vec{r})
\hat{R}(\Omega) =
\overline{{\rho}}_{-\Omega}(\vec{r}),
\end{equation}

we finally obtain  

\begin{equation} \label{hstarP}
h^{*}(\Omega) =
\langle \varphi(\vec{q}) \mid  \hat{H} 
\left[\overline{{\rho}}_{-\Omega}(\vec{r})\right]
\hat{R}(-\Omega)
\mid \varphi(\vec{q}) \rangle =
h(-\Omega).
\end{equation}

Besides the fact that with the prescription of Eq. (\ref{HAMDENJ}) one recovers for
the energy  an expression similar to the one already known for density
independent forces (Eq. (\ref{PROJENER1})), it is also important to  note that when the intrinsic wave
function is strongly deformed, and the Kamlah expansion can be used to obtain an
approximate expression for the projected  energy (cranking model), a
prescription like the one of Eq. (\ref{HAMDENJ}) yields the  correct expression for  the
angular velocity $\omega$ including the "rearrangement" term 
\cite{Valor.96,Valor.97}.

\subsection{Restriction to axially symmetric intrinsic wave functions}

For computational reasons, we restricted  ourselves to axially symmetric  ($K=0$)
configurations. In this way two of the three integrals on the Euler angles can
be performed analytically reducing by at least two orders of magnitude the
computational burden. The axially symmetric Hartree-Fock-Bogoliubov (HFB) wave
functions $\mid \varphi(q_{20})\rangle$ were obtained  in mean field
calculations  with the axial quadrupole moment  $\hat{Q}_{20}= z^{2} -
\frac{1}{2}\left(x^{2}+y^{2}\right)$ as constraining operator  
\footnote{ The definition used for the intrinsic quadrupole moment is a factor 
$\frac{1}{2}$ smaller than the usual definition for this quantity.}.
Besides the axial symmetry, we have imposed as selfconsistent symmetries the
parity, the $e^{-i \pi \hat{J}_{y}}$ symmetry  and time-reversal.

Due to the $K=0$ restriction in the  HFB states
$\mid \varphi(q_{20})\rangle$ (i.e.,  $\hat{J}_z\mid \varphi(q_{20})\rangle=0$)
we obtain

\begin{eqnarray}
\overline{\rho}_{\Omega}(\vec{r}) &=&
\frac{\langle \varphi(q_{20}) \mid \hat{\rho}(\vec{r})
\hat{R}(\Omega)
 \mid \varphi(q_{20}) \rangle}
{\langle \varphi(q_{20}) \mid \hat{R}(\Omega)\mid \varphi(q_{20}) \rangle}
=
\frac{\langle \varphi(q_{20}) \mid \hat{\rho}(\vec{r})
e^{-i \alpha \hat{J}_{z}}e^{-i \beta \hat{J}_{y}}
 \mid \varphi(q_{20}) \rangle
}
{\langle \varphi(q_{20}) \mid 
e^{-i \beta \hat{J}_{y}}
 \mid \varphi(q_{20}) \rangle
} 
\nonumber\\
&=&
\frac{\langle \varphi(q_{20}) \mid
e^{i \alpha \hat{J}_{z}}
 \hat{\rho}(\vec{r})
e^{-i \alpha \hat{J}_{z}}e^{-i \beta \hat{J}_{y}}
 \mid \varphi(q_{20}) \rangle
}
{\langle \varphi(q_{20}) \mid 
e^{-i \beta \hat{J}_{y}}
 \mid \varphi(q_{20}) \rangle
}
=
\overline{\rho}_{\beta}(\mathcal{R}^\dagger_{z}(\alpha)\vec{r})
\end{eqnarray}
where $\mathcal{R}_z(\alpha)$ is the rotation matrix along the $z$ axis. In
the same way

\begin{eqnarray} \label{simplif_1}
h(\Omega) &=&
\langle \varphi(q_{20}) \mid    
 \hat{H} \left[\overline{\rho}_{\Omega}(\vec{r}) \right]
 \hat{R}(\Omega) \mid \varphi(q_{20}) \rangle
 \nonumber\\
&=& 
\langle \varphi(q_{20}) \mid    
 \hat{H} \left[
\overline{\rho}_{\beta}(\mathcal{R}^\dagger_{z}(\alpha)\vec{r}) 
  \right]
  e^{-i \alpha \hat{J}_{z}}e^{-i \beta \hat{J}_{y}}
 \mid \varphi(q_{20}) \rangle
 \nonumber\\
&=&
\langle \varphi(q_{20}) \mid
e^{i \alpha \hat{J}_{z}}    
 \hat{H} \left[
\overline{\rho}_{\beta}(\mathcal{R}^\dagger_{z}(\alpha)\vec{r}) 
  \right]
  e^{-i \alpha \hat{J}_{z}}e^{-i \beta \hat{J}_{y}}
 \mid \varphi(q_{20}) \rangle 
 \nonumber\\
&=&
\langle \varphi(q_{20}) \mid   
 \hat{H} \left[ 
\overline{\rho}_{\beta}(\vec{r}) 
  \right]
e^{-i \beta \hat{J}_{y}}
 \mid \varphi(q_{20}) \rangle
 =h(\beta)   
\end{eqnarray}

Using the selfconsistent symmetry
$e^{-i \pi \hat{J}_{y}}$ in the 
HFB wave functions 
$\mid \varphi(q_{20})\rangle$
and the identity 
$e^{i \beta \hat{J}_{y}} = e^{-i \pi \hat{J}_{z}} 
e^{-i \beta \hat{J}_{y}} e^{i \pi \hat{J}_{z}}$ it can be easily shown
that  $\overline{\rho}_\beta (\vec{r})$ and $h(\beta)$ are real quantities
that also satisfy
$\overline{\rho}_{\beta} (\vec{r})=\overline{\rho}_{\pi-\beta} (\vec{r})$
and $h(\beta)=h(\pi-\beta)$. Using these properties and
$d_{00}^{I}(\pi - \beta)=(-)^{I}d_{00}^{I}(\beta)$ 
the integration interval on $\beta$ can be reduced to $[0,\pi/2]$ and it 
can be shown that the integrals are only different from zero when $I$ is even 
(notice that the selfconsistent symmetry $e^{-i \pi \hat{J}_{y}}$ forces
positive parity intrinsic states). Therefore, the projected energy is a well
defined quantity for even $I$ and for odd $I$ becomes an indeterminacy.
The projected energy can finally 
be written, for each $q_{20}$-configuration  in the following form

\begin{eqnarray} \label{PROJENER_AXIAL}
E^{I}(q_{20}) &=&
\Delta (I) \frac{ 
\int_{0}^{\frac{\pi}{2}} d\beta \sin (\beta) d_{00}^{I *}(\beta) 
\langle \varphi(q_{20}) \mid   
 \hat{H} \left[ 
\overline{\rho}_{\beta}(\vec{r}) 
  \right]
e^{-i \beta \hat{J}_{y}}
 \mid \varphi(q_{20}) \rangle}{
\int_{0}^{\frac{\pi}{2}} d\beta \sin (\beta) d_{00}^{I *}(\beta)
\langle \varphi(q_{20}) \mid   
e^{-i \beta \hat{J}_{y}}
 \mid \varphi(q_{20}) \rangle}
\nonumber\\
&=&
\Delta (I) \frac{ 
\int_{0}^{\frac{\pi}{2}} d\beta \sin (\beta) d_{00}^{I *}(\beta) 
h(\beta) 
}{
\int_{0}^{\frac{\pi}{2}} d\beta \sin (\beta) d_{00}^{I *}(\beta) n(\beta)
}
\end{eqnarray}
where $\Delta (I)=\frac{1}{2} \left(1+(-)^ I\right)$ has been introduced to
recall that for odd $I$ the projected energy is an indeterminacy. In the above
expression for the energy, the density to be used in the density dependent part
of the interaction is given by 

\begin{eqnarray} \label{rho_axial}
\overline{\rho}_{\beta}(\vec{r})
=
\frac{\langle \varphi(q_{20}) \mid \hat{\rho}(\vec{r})
e^{-i \beta \hat{J}_{y}}  
 \mid \varphi(q_{20}) \rangle}
{\langle \varphi(q_{20}) \mid 
e^{-i \beta \hat{J}_{y}}  
\mid \varphi(q_{20}) \rangle}
\end{eqnarray}

Finally we would like to mention that in order to account for the fact that 
the mean value of the particle's number operator usually differs from the 
nucleus' proton and neutron numbers, we followed the usual recipe (see for example
\cite{HARA_RING,GCM})
 and 
replaced $h(\beta)$ by $h^{'}(\beta)=h(\beta)
-\lambda_{Z}\left( Z(\beta) -Z_{0}\right)- \lambda_{N}\left( N(\beta) -N_{0}\right)$  
where $Z(\beta)=\langle \varphi(q_{20}) \mid  \hat{Z} 
e^{-i \beta \hat{J}_{y}} 
 \mid \varphi(q_{20}) \rangle$,  $N(\beta)=\langle \varphi(q_{20}) \mid  \hat{N} 
e^{-i \beta \hat{J}_{y}} 
 \mid \varphi(q_{20}) \rangle$ and  
$\lambda_{Z}$ and $\lambda_{N}$ are chemical potentials for protons 
and neutrons, respectively.

\subsection{Angular momentum projected configuration mixing 
with density 
dependent interactions.}
\label{GCMSECTION}

Once we have the angular momentum projected potential energy surfaces
(AMPPES), the next step is to carry out  
configuration mixing (AMPGCM).  In this case the ansatz for the 
wave function is 

\begin{eqnarray} \label{GCM_ANSATZ}
\mid  \Phi_{IM} (\sigma) \rangle =
\int d \vec{q} f^{I,\sigma}(\vec{q}) 
\mid \Psi_{IM} (\vec{q})\rangle
\end{eqnarray}
where  the wave functions $\mid \Psi_{IM} (\vec{q})\rangle$ are given by
Eq. (\ref{PROJWF}). The amplitudes 
$f^{I,\sigma}(\vec{q})$ are solutions of the Hill-Wheller
(HW) equation \cite{HILL_W,ref_RING}. As it was mentioned 
before, for the moment, we restricted ourselves to axially 
symmetric ($K=0$) configurations and in this case the amplitudes
$f^{I,\sigma}(q_{20})$ are found by solving the  
HW equation 

\begin{equation} \label{HW_EQ}
\int dq_{20}^{'}  \left(\mathcal{H}^{I}(q_{20},q_{20}^{'}) 
- E^{I,\sigma} \mathcal{N}^{I}(q_{20},q_{20}^{'}) \right)
f^{I,\sigma}(q_{20}^{'}) = 0
\end{equation}
with the kernels $\mathcal{H}^{I}(q_{20},q_{20}^{'}) $
and $\mathcal{N}^{I}(q_{20},q_{20}^{'})$
defined, for even spin values,  as 

\begin{equation} \label{hamiltonian_kernel}
\mathcal{H}^{I}(q_{20},q_{20}^{'})=
\left(2I+ 1\right)
\int_{0}^{\frac{\pi}{2}}
 d \beta  \sin (\beta) d_{00}^{I *}(\beta) 
\langle \varphi(q_{20}) \mid \hat{H} \left[\overline{{\rho}}_{\beta}
^{GCM}
(\vec{r})\right]
e^{-i \beta \hat{J}_{y}}
\mid  \varphi(q_{20}^{'})
\rangle
\end{equation}
and
\begin{equation}
\mathcal{N}^{I}(q_{20},q_{20}^{'})=
\left(2I+ 1\right)
\int_{0}^{\frac{\pi}{2}}
 d \beta  \sin (\beta) d_{00}^{I *}(\beta) 
\langle \varphi(q_{20}) \mid 
e^{-i \beta \hat{J}_{y}}
\mid  \varphi(q_{20}^{'})
\rangle
\end{equation}

The previous results for $\mathcal{H}^{I}(q_{20},q_{20}^{'})$ and 
$\mathcal{N}^{I}(q_{20},q_{20}^{'})$ can be found along the same 
lines described in the previous section. In 
Eq. (\ref{hamiltonian_kernel}) the density 
$\overline{{\rho}}_{\beta}^{GCM}(\vec{r})$ is given by

\begin{equation}  
\overline{\rho}_{\beta}^{GCM}(\vec{r}) = 
\frac{\langle \varphi(q_{20}) \mid \hat{\rho} e^{-i \beta \hat{J}_{y}}
\mid \varphi(q_{20}^{'})  \rangle}
{\langle \varphi(q_{20}) \mid  e^{-i \beta \hat{J}_{y}}
\mid \varphi(q_{20}^{'})  \rangle
}
\end{equation}
which is the generalization of Eq. (\ref{rho_axial}) 
for the density dependent part of the interaction
in the framework of the configuration mixing calculation. As in the 
case of angular momentum projection  we also replaced 
$\hat{H} \left(\overline{\rho}_{\beta}^{GCM}(\vec{r}) \right)$
by 
$\hat{H} \left(\overline{\rho}_{\beta}^{GCM}(\vec{r}) \right)
- \lambda_{Z}\left( \hat{Z} -Z_{0}\right) - \lambda_{N}\left( \hat{N} -N_{0}\right)$
where $\lambda_{Z}$ and $\lambda_{N}$ are chemical potentials for protons and 
neutrons, respectively.

The first step \cite{ref_RING} in the solution of the HW equation Eq. (\ref{HW_EQ}) is to 
diagonalize the norm kernel 
$$
\int d q_{20}^{'} \mathcal{N}^{I}(q_{20},q_{20}^{'}) u_k^I(q_{20}^{'})=
n_k^I u_k^I (q_{20}).
$$
The norm eigenvalues $n_{k}^{I}$
with zero values are subsequently  
removed (i.e, linearly dependent states are removed from the basis) 
\footnote{In practical computations eigenvalues smaller than a given threshold 
$\epsilon$ should be removed to ensure the numerical stability 
of the solution of the HW equation.}
for a 
proper definition of the collective image of the kernel
$\mathcal{H}^{I}(q_{20},q_{20}^{'})$ which is given by

\begin{eqnarray}
H^{{\Im}_{C}}_{kk^{'}}(I)
= 
\frac{1}{\sqrt{n_{k}^{I}} \sqrt{n_{{k}^{'}}^{I}}}
\int dq_{20} \int dq_{20}^{'}  u_{k}^{I *}(q_{20}) \mathcal{H}^{I}(q_{20},q_{20}^{'})
u_{k^{'}}^{I}(q_{20}^{'})
\end{eqnarray}
and the diagonalization in the collective subspace $\Im_{C}$ 

\begin{eqnarray} \label{price}
\sum_{k k^{'}} H^{{\Im}_{C}}_{kk^{'}}(I) g_{k^{'}}^{I, \sigma} =
E^{I,\sigma} g_{k}^{I, \sigma}
\end{eqnarray}
gives us the energy $E^{I,\sigma}$ for each spin value not only for  
the ground state ($\sigma =1$) but also  for excited states 
($\sigma =2,3,4, \dots$) that, with the considered set of generating wave 
functions, could correspond to states with a different deformation from the one 
of the 
ground state and/or to quadrupole vibrational states.

Using the eigenfunctions $u_{k}^{I}(q_{20})$ of the norm 
kernel  $\mathcal{N}^{I}(q_{20},q_{20}^{'})$ and the amplitudes
$g_{k}^{I, \sigma}$ (see Eq. (\ref{price})) we can compute \cite{ref_RING}
the amplitudes $f^{I,\sigma}(q_{20})$

\begin{eqnarray} \label{FFF}
f^{I,\sigma}(q_{20}) = 
\sum_{k} \frac{g_{k}^{I, \sigma}}{\sqrt{n_{k}^{I}}}
u_{k}^{I}(q_{20}) 
\end{eqnarray}
and the so called collective wave functions 
$G^{I,\sigma}(q_{20})$ 

\begin{eqnarray}
G^{I,\sigma}(q_{20}) =
\sum_{k} g_{k}^{I, \sigma} u_{k}^{I}(q_{20}).
\end{eqnarray}
These collective wave functions are orthonormal and therefore their module 
squared can be interpreted as a probability amplitude.

Finally, from the knowledge of the amplitudes 
$f^{I,\sigma}(q_{20})$, we can compute the reduced
$B(E2)$ transition probabilities
and the spectroscopic quadrupole moments $Q^{spec}(I,\sigma)$. This 
is one of the main motivations for 
carrying out a configuration mixing calculation 
of angular momentum projected wave functions 
in the case of Gogny 
\cite{ref_ff,ref_gg,ref_ii,ref_32S,N=28}
and Skyrme \cite{ANGEL_24Mg,ref_ee}
forces, since both interactions allow the use of full configuration spaces 
and then one is able to compute transition probabilities
and spectroscopic quadrupole moments
 without effective
charges.  
In the framework of the AMPGCM the $B(E2)$ transition probability
between the  states $(I_{i}, {\sigma}_{i} )$ and 
$(I_{f},{\sigma}_{f})$
is expressed as
(see Appendix A for more details)
 
\begin{eqnarray} \label{BE2_GCM_AM}
&B&(E2,{I_{i}} {{\sigma}_{i}} \rightarrow {I_{f}} {{\sigma}_{f}})
=
\frac{e^{2}}{2I_{i}+1} \\ \nonumber
& \times &\left | 
\int dq_{20,i} dq_{20,f}  f^{I_{f},{\sigma}_{f} *}(q_{20,f}) 
\langle I_{f} q_{20,f} \mid \mid \hat{Q}_{2} \mid \mid
I_{i} q_{20,i} \rangle
f^{I_{i},{\sigma}_{i} }(q_{20,i})
\right |^{2}
\end{eqnarray}
and the spectroscopic quadrupole 
moment for the state $(I \ge 2,\sigma)$ 
is given by

\begin{eqnarray} \label{ESPECT_GCM_AM}
&Q&^{spec}(I,\sigma) = e  \sqrt{\frac{16 \pi}{5}}
\left ( \begin{array} {ccc}
        I & 2 &  I  \\
        I & 0 & -I
        \end{array}  \right) \\ \nonumber
&\times& \int dq_{20,i} dq_{20,f}  f^{I, \sigma *}(q_{20,i}) 
\langle I q_{20,i} \mid \mid \hat{Q}_{2} \mid \mid
I  q_{20,f} \rangle
f^{I , \sigma}(q_{20,f}) 
\end{eqnarray}

\subsection{Details on the calculation}

The intrinsic wave functions $\mid \varphi (q_{20}) \rangle$ were obtained as
the solutions of the constrained axially symmetric HFB equations with the
constraint on  the mass quadrupole moment. The HFB creation and annihilation
operators were expanded in an axially symmetric Harmonic Oscillator (HO) basis
including  ten major shells (i.e., $N_{Shell}=10$). The two oscillator length parameters of the basis
$b_\perp$ and $b_z$ were chosen to be always equal in order to keep the basis
closed  under rotations \cite{Robledo.94,Egido.93} (this is
also the reason to include full HO major shells in the basis). The same
oscillator length was used for all the quadrupole deformations considered in
order to avoid completeness problems in the GCM
calculations \cite{Robledo.94}.

In the HFB calculations the two body kinetic energy correction was fully taken
into account both in the energy and in the minimization procedure. This term is
specially important in the nuclei considered due to their small mass number.
Concerning the Coulomb interaction, both the exchange and pairing parts were
not taken into account as they increase the computational burden by at least an
order of magnitude. However, we computed the Coulomb exchange energy in the
Slater approximation and added this quantity at the end of the calculations in
a perturbative fashion.

As self consistent symmetries we kept, in addition to the axial symmetry, 
the parity (no octupole mixing is allowed), the $e^{-i\pi J_y}$ symmetry and 
time reversal. 

For the computation of the matrix elements of the rotation operator in the 
HO basis we have used the results of \cite{ROT_MATRIX}. 

For the evaluation of overlaps of one or two body operators between different HFB
wave functions we have used 
the extended Wick's theorem \cite{Balian69}. In order to determine the sign of
the norm overlaps we have followed the procedure proposed in Ref.
\cite{Neergard83}.

\section{DISCUSSION OF THE RESULTS.} 
 
\subsection{Mean field approximation for Magnesium isotopes.}
\label{mean-field}

\begin{figure}
\begin{center}
\includegraphics[angle=0,width=6cm]{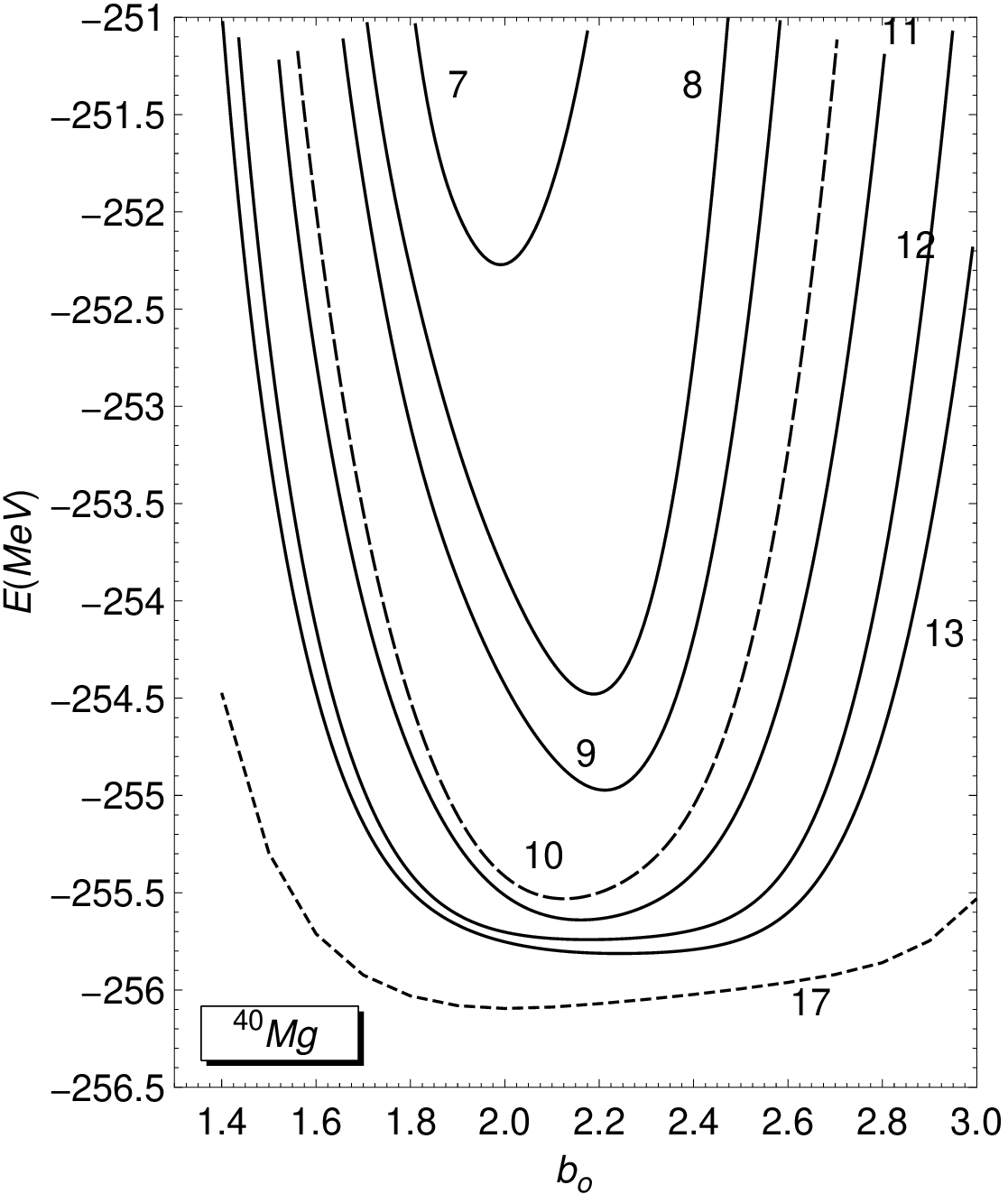}\includegraphics[angle=0,width=6cm]{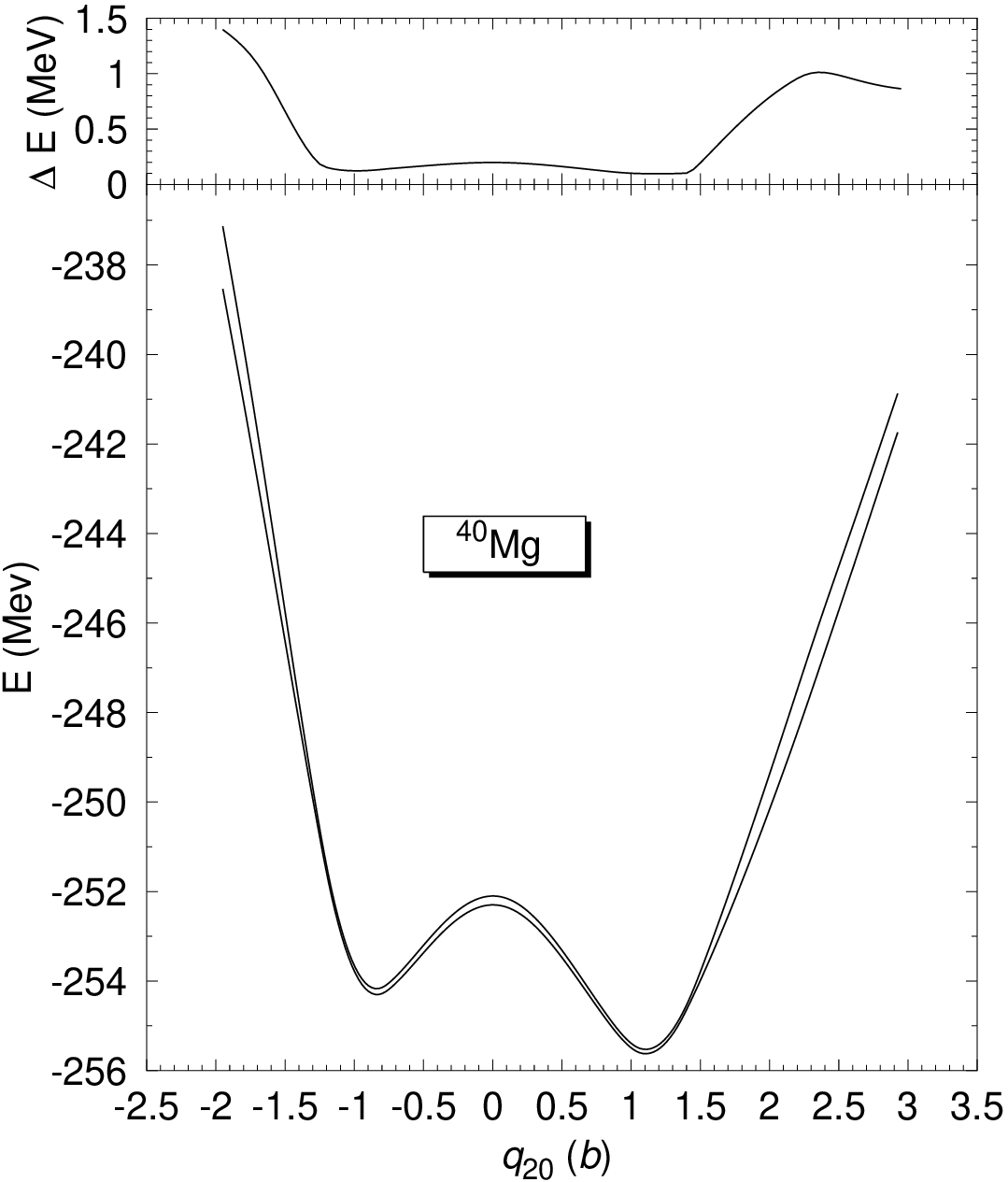}

\caption{On the left hand side, the ground state HFB energies for \nuc{40}{Mg} 
are plotted as functions of the HO
length $b_{o}$ ($b_\perp=b_z=b_o$) for  the  $N_{Shell}=7,8,9,10,11,12,13$ and 17 bases. 
The curve corresponding to the basis used in the present work 
($N_{Shell}=10$) is plotted as a dashed line. On the right hand side, lower
panel, the 
HFB energies computed with $b_{o}=2.1$ fm and $N_{Shell}=10$ and 11 are plotted as
a function of the quadrupole moment. In the upper panel, the energy difference
between the $N_{Shell}=10$ and $N_{Shell}=11$ calculations is plotted as a
function of the quadrupole moment.}

\label{Fig_4_art}
\end{center} 
\end{figure}

A few words concerning the convergence of our calculations with the size of
the HO basis are in order here specially taking into account that we have to
deal with  the dripline nucleus \nuc{40}{Mg}. One should note that due to  the proximity of the two neutron
dripline, the full HFB approximation must be used \cite{Doba.84,ref_i}
and absolute convergence for the binding energy can only  be found for HO basis
with a very
large number of shells. At the mean field  level such a hard task is still
feasible. Even in the case we were interested  in a single $q_{20}$
configuration, angular momentum projected calculations with  very large
$N_{Shell}$ can also be performed. The main reason why we can not  consider so
large $N_{Shell}$ in the present study is obvious: the enormous amount
(typically of the order of one thousand) of angular momentum projected
hamiltonian overlaps to be computed in the configuration mixing calculation
make the problem intractable. In \nuc{40}{Mg}  (using the   $q_{20}$ range
$-2.0b \le q_{20} \le 3.0b$, the  mesh  $\Delta q_{20}=0.1b$ and 16 points in the
angular momentum projection) if   the basis is increased 
from $N_{Shell}=10$ to $N_{Shell}=11$  the computational time required to
evaluate all the projected  kernels increases by a factor  of 10 (from 3
days to one month in a typical  workstation). Here it is  worth to remark that
besides the technical difficulties in the  configuration mixing calculation,
our force is a finite range force  and its mathematical structure leads to very
complicated matrix elements whose evaluation is very time consuming. 

On the
other hand, as the absolute value of the  binding energy does not affect very
much the collective motion (it is only affected by the shape of the energy landscape) we
can  select a reasonable value of $N_{Shell}$ for which the energy landscape
is well converged (i.e. its shape remains almost unchanged in the regions of
physical significance when the value of $N_{Shell}$ is increased). As we will
see in the next paragraph $N_{Shell}=10$ is more than enough for the present
study.

On the left hand side of Fig. \ref{Fig_4_art} we show the ground state  HFB
energy for \nuc{40}{Mg} as a function of the oscillator length $b_{o}$ 
($b_\perp=b_z=b_o$) for
different values of $N_{Shell}$ (the Coulomb exchange energy is not included).
As expected the curves  become more and more flat for increasing values of 
$N_{Shell}$. Already the $N_{Shell}=17$ basis can be considered as a
reasonable  approximation for an infinite basis in this nucleus as the
dependence of the energy on the oscillator length is very weak for a wide range
of $b_o$ values. For  $N_{Shell} =10$ a minimum in the energy  curve is
obtained for $b_{o}=2.1$. Using the minima of the corresponding energy curves 
for $N_{Shell}=10$  and $N_{Shell}=17$ we get   $E_{N_{Shell}=10} -
E_{N_{Shell}=17} = 568.37$ keV. In \nuc{38}{Mg}, the same  analysis has been
carried out and the overestimation of the energy is 543.29 keV. As a
consequence,  for the two neutron separation energy  $S_{2N}$ it is obtained
that  $\mid S_{2N,N_{Shell}=10}(^{40}Mg)-S_{2N,N_{Shell}=17}(^{40}Mg)\mid =
25.12 $ keV.  Also a very good agreement  is found, in \nuc{40}{Mg},  between 
our values for the proton and neutron $\beta_{2}$ deformation parameters
$({\beta}_{2}^{Z},{\beta}_{2}^{N})=(0.38,0.31)$ and the ones (0.35, 0.28) of the 
calculation in coordinate space of Ref. \cite{ref_cc}. This can be understood
from the fact that  our basis roughly corresponds to  $R_{box} \approx 9.4 fm$
while it was found in  Ref. \cite{ref_cc} that the deformation parameters
remain practically unchanged for $R_{box} \ge 7 fm$ in \nuc{40}{Mg}.  

On the right hand side of Fig. \ref{Fig_4_art} we show, in the lower panel, 
the energy landscapes of \nuc{40}{Mg} as a function of the quadrupole moment
for the calculations with $N_{Shell}=10$ and $N_{Shell}=11$. In the upper panel
we have represented the energy difference between both calculations. From this
plots we observe how in the region between -1.5 b and 1.5 b the energy
landscape does not change much when the basis is increased. As we will see in
another  section this range of quadrupole deformation is the one where the
collective dynamics is concentrated and therefore it is not expected to find
significant differences between the calculations with $N_{Shell}=10$ and
$N_{Shell}=11$.

 Since the main interest of the present study  is focussed on  physical
quantities like $S_{2N}$ values, rotational  energy corrections, excitation
energies, etc.,  and these quantities  do not change very much with the number
of HO shells, we conclude  that our mean field results with $N_{Shell}=10$  can
be regarded as a reasonable choice. Even  more, the zero point rotational
energy correction  $E_{ROT}^{I=0}= E_{HFB}-E^{I=0}$ is $ \approx$  3 MeV for
the ground state configuration in \nuc{40}{Mg}, i.e, the zero point rotational 
energy correction is 5.3 times larger than  the relative energy difference 
$E_{N_{Shell}=10} - E_{N_{Shell}=17}$, and also remains practically unchanged
for increasing values of $N_{Shell}$.

\begin{figure}
\begin{center}
\includegraphics[angle=0,width=14cm]{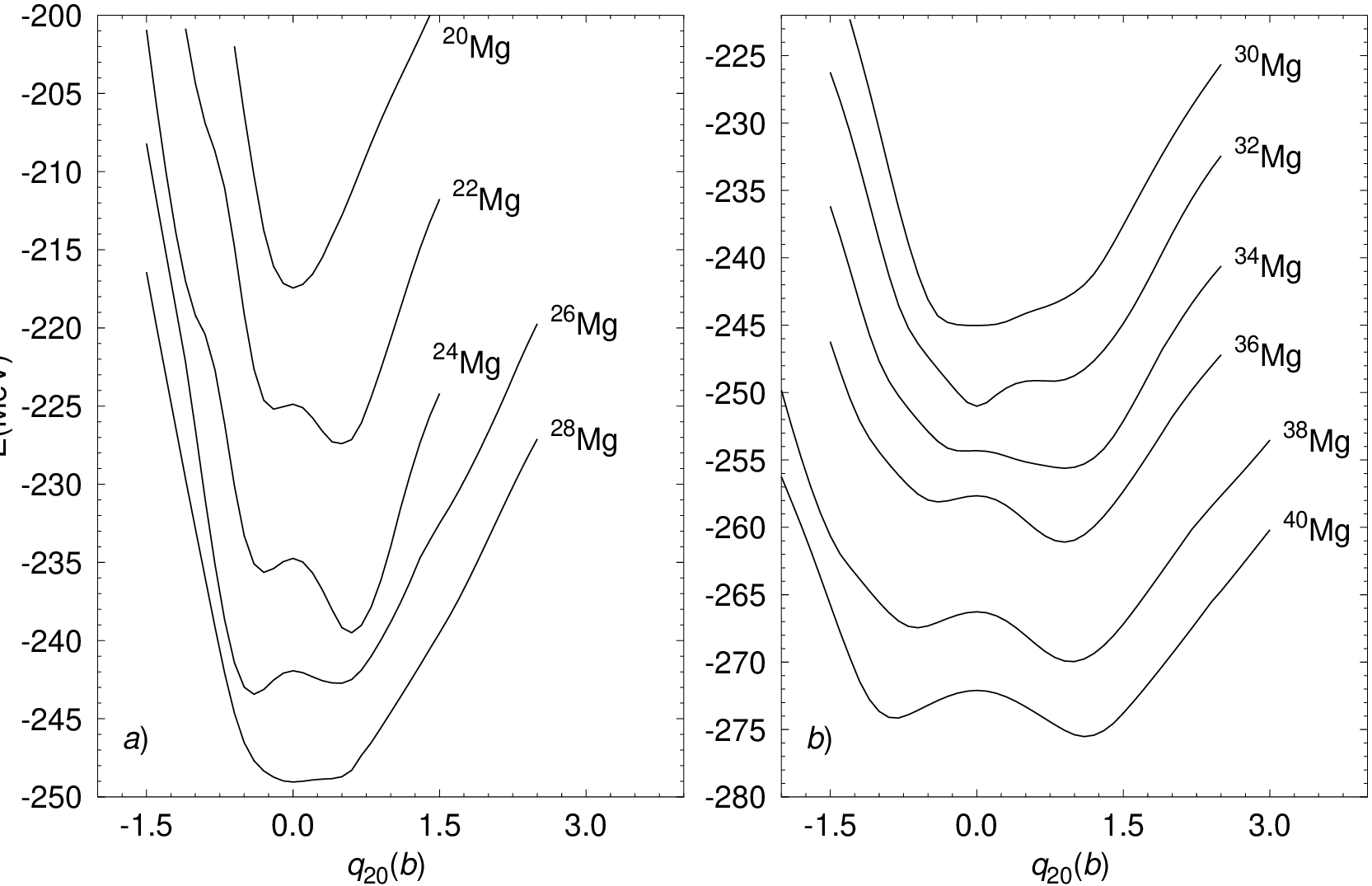}

\caption{Mean field potential energy surfaces for Magnesium isotopes  as
functions of the axially symmetric quadrupole moment. These curves have been 
shifted to accommodate them in a single plot. The corresponding energy shifts 
in  panel a) are  -85, -65, -50, -35, -25 MeV  for \nuc{20-28}{Mg}, while in
panel b) they are -10, -7, -7, -8, -15, -20 MeV for \nuc{30-40}{Mg}.}

\label{Fig_1_art}
\end{center} 
\end{figure}

In Fig. \ref{Fig_1_art} we show  the mean field potential energy surfaces
(MFPES) as a function of the axially symmetric quadrupole moment $q_{20}$ for
the even-even Magnesium isotopes \nuc{20-40}{Mg}. The MFPES shown do not
include the Coulomb exchange energy and they have been shifted to accommodate
them in a single plot (see caption). As one can see, with the exception of
\nuc{20}{Mg}, the  MFPES are very flat around the corresponding minima 
indicating that further correlations could, in some cases, change the
conclusions  obtained at the mean field level.

The nucleus \nuc{20}{Mg} shows a well defined spherical minimum which is a
consequence of the N=8 neutron shell closure. On the other hand, both
\nuc{22}{Mg} and \nuc{24}{Mg} are prolate deformed in their ground states. In 
\nuc{22}{Mg} the ground state corresponds to $q_{20}=0.5b({\beta}_{2} = 0.40)$
and an oblate local minimum also appears at $q_{20}=-0.2b$ (${\beta}_{2} =
-0.17$) with an excitation energy of 1.65 MeV. In the case of \nuc{24}{Mg} the
ground state  corresponds to $q_{20}=0.6b$ (${\beta}_{2} = 0.43$) and another
local minimum is  found at $q_{20}=-0.3b$ (${\beta}_{2} = -0.22$) with an
excitation energy of  3.86 MeV. The only  oblate isotope in this chain is the
nucleus \nuc{26}{Mg} whose ground  state is located at $q_{20}=-0.4b$
(${\beta}_{2} = -0.26$). A prolate isomeric state is also found at 
$q_{20}=0.5b$  (${\beta}_{2} = 0.32$) with an excitation energy of  707 keV
with respect to the oblate ground state. On the other hand, the nuclei
\nuc{28}{Mg}, \nuc{30}{Mg} and \nuc{32}{Mg} show spherical ground states. The 
MFPES of both \nuc{28,30}{Mg} are  particularly flat around their  spherical
ground states. In the nucleus \nuc{32}{Mg} we obtain a  prolate shoulder at
$q_{20}=0.8b$ (${\beta}_{2} = 0.36$) in the MFPES with 1.86 MeV of excitation
energy with  respect to the spherical ground state.

From \nuc{34}{Mg} to the dripline isotope \nuc{40}{Mg}, prolate deformed ground
states are found. The ground state deformations  are  $q_{20}=0.9b $ $ 
({\beta}_{2} = 0.36)$, $q_{20} =0.9b $ $ ({\beta}_{2} = 0.33)$, 
$q_{20}=1.0b $ $  ({\beta}_{2} = 0.33)$ and $q_{20}=1.1b $ $ ({\beta}_{2}
= 0.33)$ in \nuc{34}{Mg}, \nuc{36}{Mg}, \nuc{38}{Mg} and \nuc{40}{Mg},
respectively.  Besides the fact that prolate configurations become dominant in
all these nuclei, one should note that close-lying oblate isomeric states are
found in the MFPES of  \nuc{36}{Mg}, \nuc{38}{Mg} and \nuc{40}{Mg}  at
$q_{20}=-0.4b$ $({\beta}_{2} = -0.15)$, $q_{20}=-0.6b$ $ ({\beta}_{2}
= -0.20)$ and  $q_{20}=-0.8b$ $ ({\beta}_{2} = -0.25)$ with
excitation energies of 2.98 MeV, 2.51 MeV and  1.38 MeV with respect to 
prolate ground states.

\begin{figure}
\begin{center}
\includegraphics[angle=0,width=13cm]{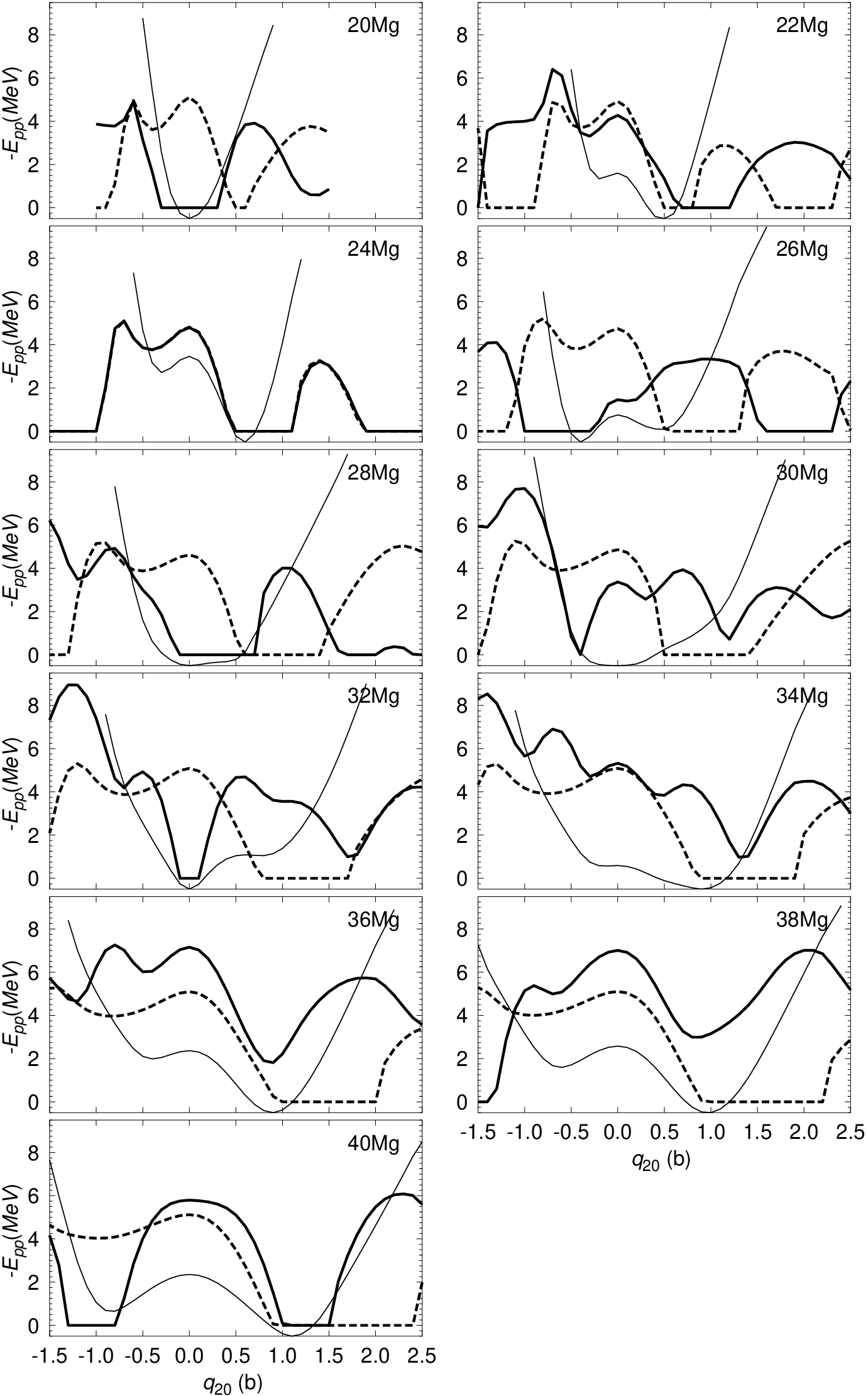}
\caption{In each panel, proton (thick dashed lines) and neutron (thick full lines) 
particle-particle correlation energies $-E_{pp}$ are depicted as a function of the
quadrupole moment. Also the HFB energy curves  are  plotted as 
thin full lines.}
\label{Fig_3_art}
\end{center} 
\end{figure}

In  Fig. \ref{Fig_3_art} the proton and neutron  particle-particle 
correlation  energies ( defined as $-E_{pp} = \frac{1}{2} Tr\left( \Delta
{\kappa}^{*}\right)$) are plotted as a function of the quadrupole deformation
for all the isotopes considered. The evolution  of the  particle-particle 
correlation  energies is well correlated with the structures found in the 
MFPES. Non-zero proton pairing correlations are found in all the  spherical or
oblate minima. In addition, sizeable  neutron pairing correlations are found in
\nuc{22,24}{Mg} and  in  \nuc{36,38,40}{Mg} for the spherical and oblate
minima. Vanishing  proton pairing correlations are found  in the prolate side
starting at $q_{20}=0.5$b in all the isotopes. The range of quadrupole moments
for which the proton pairing correlations vanish increases with the neutron
number. An expected result is that neutron  pairing correlations vanish in the
mean field spherical ground states of both  \nuc{20}{Mg} and \nuc{32}{Mg} as a
consequence of  the  N=8 and N=20 shell closures. However, tangible  neutron
pairing  correlations are found at the spherical configuration of \nuc{40}{Mg} 
pointing to the erosion of the N=28 spherical shell closure already at the mean
field level. On the other hand, the lowering of the level density around  the
ground state of this  nucleus ($q_{20} \approx 1b$) leads to vanishing proton
and neutron pairing correlations.  This agrees  with the results of 
\cite{Cottle-Kemper,ref_bbb,ref_hh} and  \cite{N=28} which suggest that while
the N=28 spherical shell closure  disappears for some neutron rich nuclei a new
deformed shell closure emerges on them.

It should  be stressed here  that the unphysical collapse of pairing
correlations, which is clearly visible from the results of Fig.
\ref{Fig_3_art}, is one of the main drawbacks of our  calculations at the
present stage.  One should also consider  the dynamical pairing correlations in
these nuclei and their  coupling with the quadrupole degree of freedom in the
scope of a theory  beyond the mean field in order to treat in an equal footing
both short and long range correlations.

\begin{figure}
\begin{center}
\includegraphics[angle=0,width=10cm]{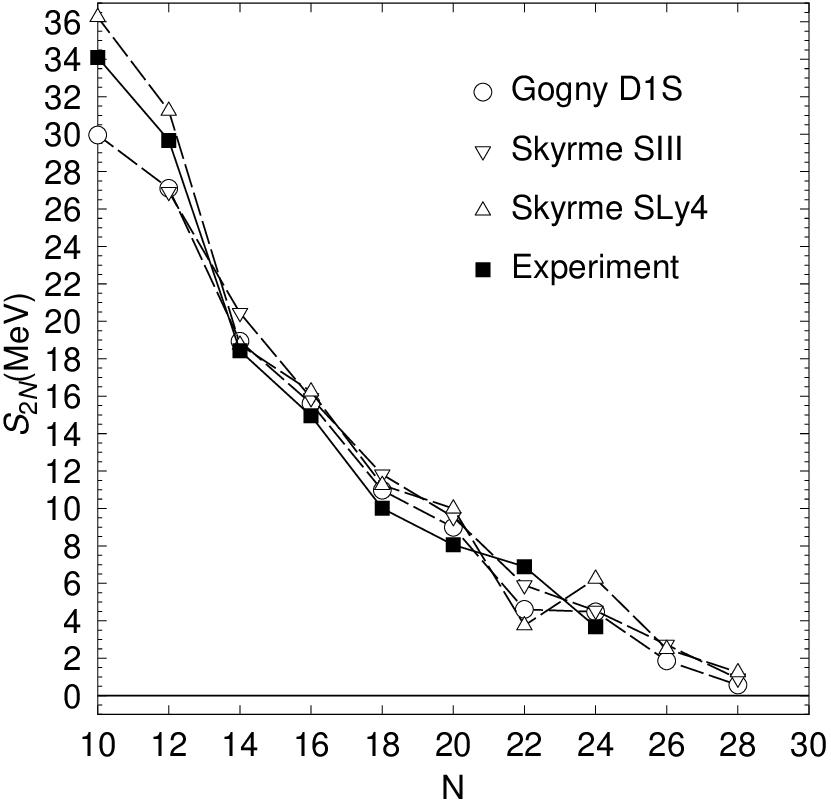}
\caption{The two neutron separation energies obtained in the present calculations 
are compared with the experimental data taken from Ref. \cite{Audi.95} and 
with the mean field results obtained  with the Skyrme interactions SIII \cite{ref_cc}
and  SLy4 \cite{ref_dd}.}
\label{Fig_6_art}
\end{center} 
\end{figure}

Using the absolute minima of the MFPES we have computed the two neutron
separation energies  $S_{2N}=E_{MF}(Z,N-2)-E_{MF}(Z,N)$ and the results are
compared in Fig. \ref{Fig_6_art}  with the ones previously obtained in the
framework of the mean field  approximation with the Skyrme force  SIII
\cite{ref_cc}  and  SLy4  \cite{ref_dd}   and with the available experimental
values \cite{Audi.95}. Our results agree quite well with those of SIII in all
the nuclei studied. In the case of SLy4 the agreement is also rather good
except for \nuc{22,24}{Mg} and \nuc{36}{Mg} where the SLy4 results are bigger
than ours. The three interactions predict that the nucleus \nuc{40}{Mg} is the
last bound isotope of the chain. Also the three interactions predict a dip  in
the $S_{2N}$ of \nuc{34}{Mg} and therefore fail  to reproduce the peak observed
experimentally. The observed peak is usually  explained as a consequence of a
deformed ground state in \nuc{32}{Mg} which has a lower energy than the
spherical configuration. As we will see later, the wrong behavior of the
$S_{2N}$ of \nuc{34}{Mg} at the mean field level is cured when the extra
correlation energy (coming from the  interplay of the quadrupole fluctuations
and the zero point rotational  energy corrections)  leading to a deformed
ground state in \nuc{32}{Mg} is considered. Finally, another significant
failure of both the Gogny and Skyrme SIII interactions is that of the predicted
values of the two neutron separation energies of \nuc{22,24}{Mg}. As we will
see later, considering the zero point rotational energy correction will improve
that situation.

\begin{figure}
\begin{center}
\includegraphics[angle=-90,width=13cm]{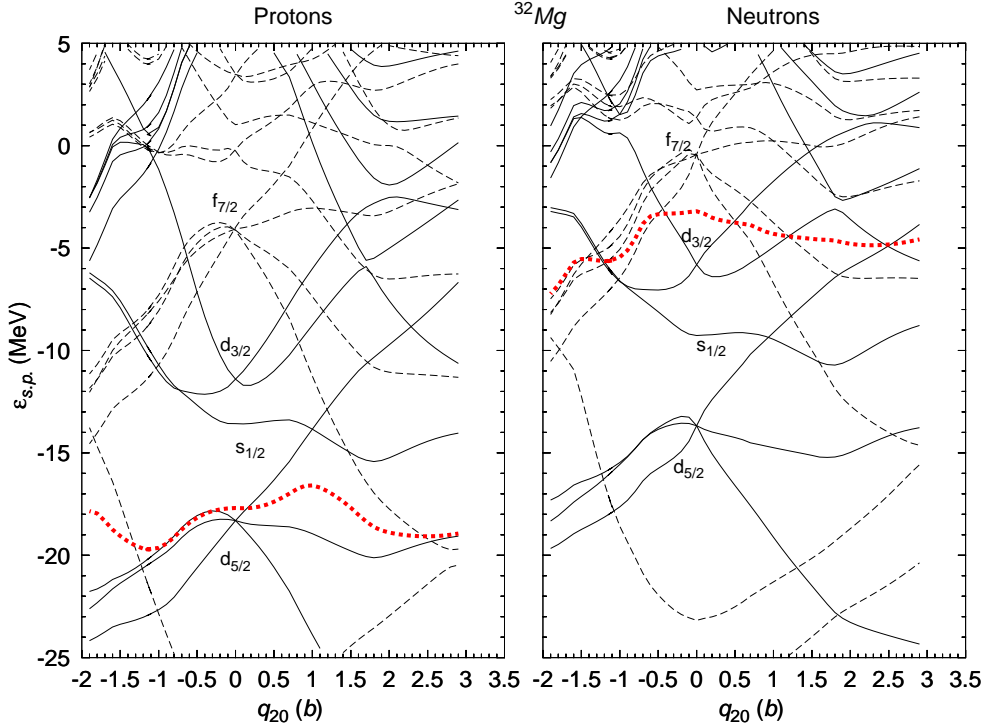}
\caption{Proton and neutron single particle diagrams as a function 
of the quadrupole deformation for the nucleus \nuc{32}{Mg}. Full (dashed)
lines correspond to positive (negative) parity levels. The Fermi 
levels are represented as thick dotted lines. }
\label{Fig_spe_32Mg}
\end{center} 
\end{figure}

In Fig. \ref{Fig_spe_32Mg} we show the single particle energies  for protons
and neutrons in \nuc{32}{Mg}. As in our mean field  calculations we   solve the
full HFB equations, the only quantities  that can be properly defined are the
quasiparticle  energies. However, in order to have the usual Nilsson like
diagrams we have chosen to plot the eigenvalues of the mean field Hartree-Fock
hamiltonian  $h=t+\Gamma$ as a function of the axially symmetric quadrupole
deformation.  Looking at the neutron single particle energies we observe how at
$q_{20}=0.8b$ a couple of $f_{7/2}$ orbitals cross the Fermi surface and become
occupied. The occupancy of those orbitals leads to the appearance of the
shoulder seen in the MFPES of \nuc{32}{Mg}. It is also interesting to notice
that the almost full  occupancy of the $d_{5/2}$ proton orbital favors oblate
shapes. In the same way, the full occupancy of the $d_{5/2}$ neutron orbital 
in  \nuc{26}{Mg} favors an oblate ground state for this nucleus.

\subsection{Correlations beyond the mean field : Angular momentum 
projection and configuration mixing.}

\begin{figure}
\begin{center}
\includegraphics[angle=0,width=14cm]{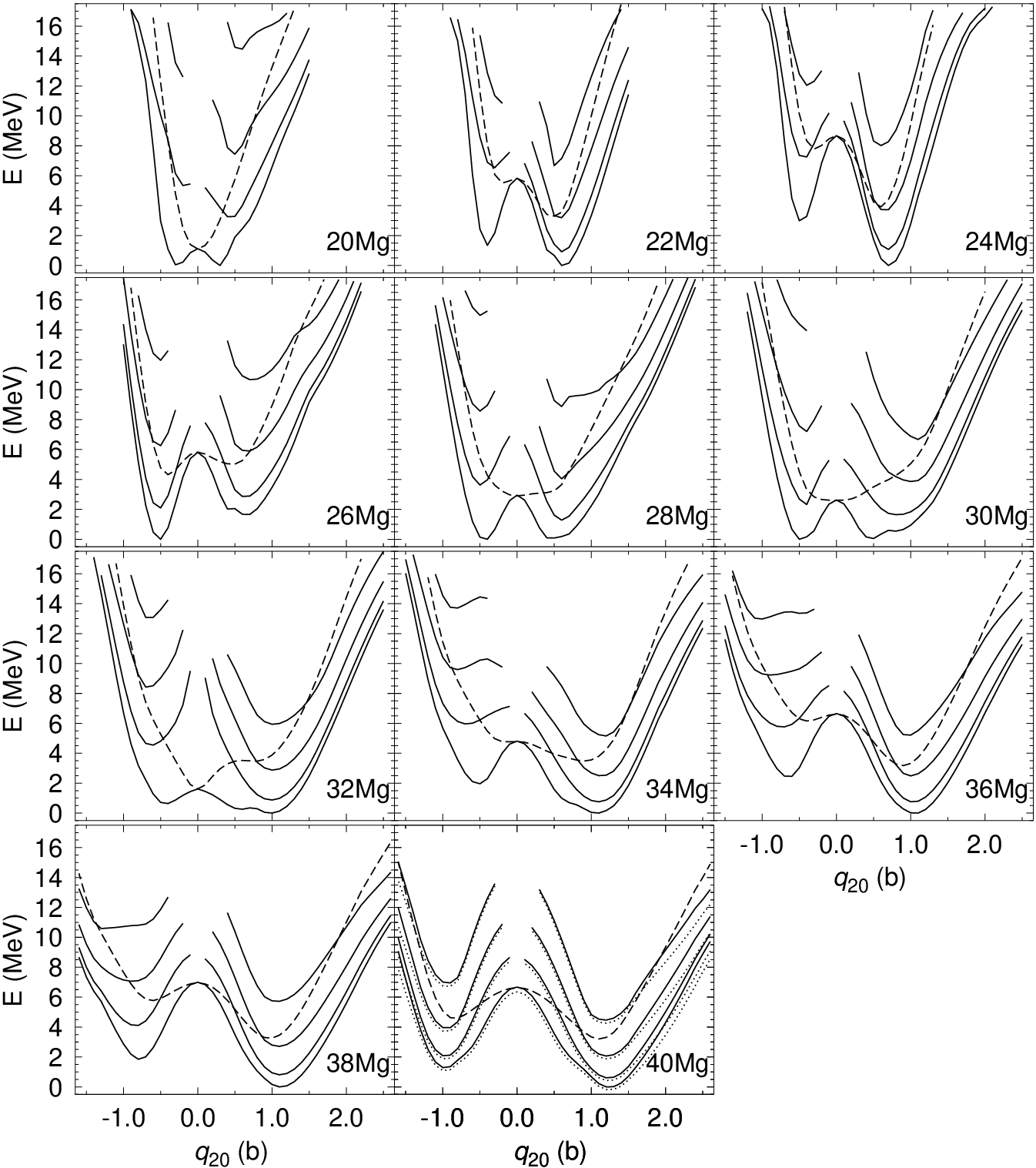}

\caption{Angular momentum projected potential energy surfaces (full lines)  for
the  nuclei \nuc{20-40}{Mg} and for the spin values
$I^{\pi}=0^{+},2^{+},4^{+},6^{+}$ as functions  of the axially symmetric
quadrupole moment $q_{20}$. The mean field potential energy surfaces are also
plotted as dashed lines. In the nucleus \nuc{40}{Mg} we have also included
(dotted lines) the
projected results corresponding to the calculation with $N_{shell}=11$.}

\label{Fig_8_art}
\end{center} 
\end{figure}

The main outcome of our angular momentum projected calculations is  presented
in Fig. \ref{Fig_8_art}, where the angular momentum  projected potential energy
surfaces (AMPPES) are shown for the nuclei \nuc{20-40}{Mg}. The corresponding
mean field energy landscapes (dashed curves) are also included for comparison.

\begin{figure}
\begin{center}
\includegraphics[angle=0,width=8cm]{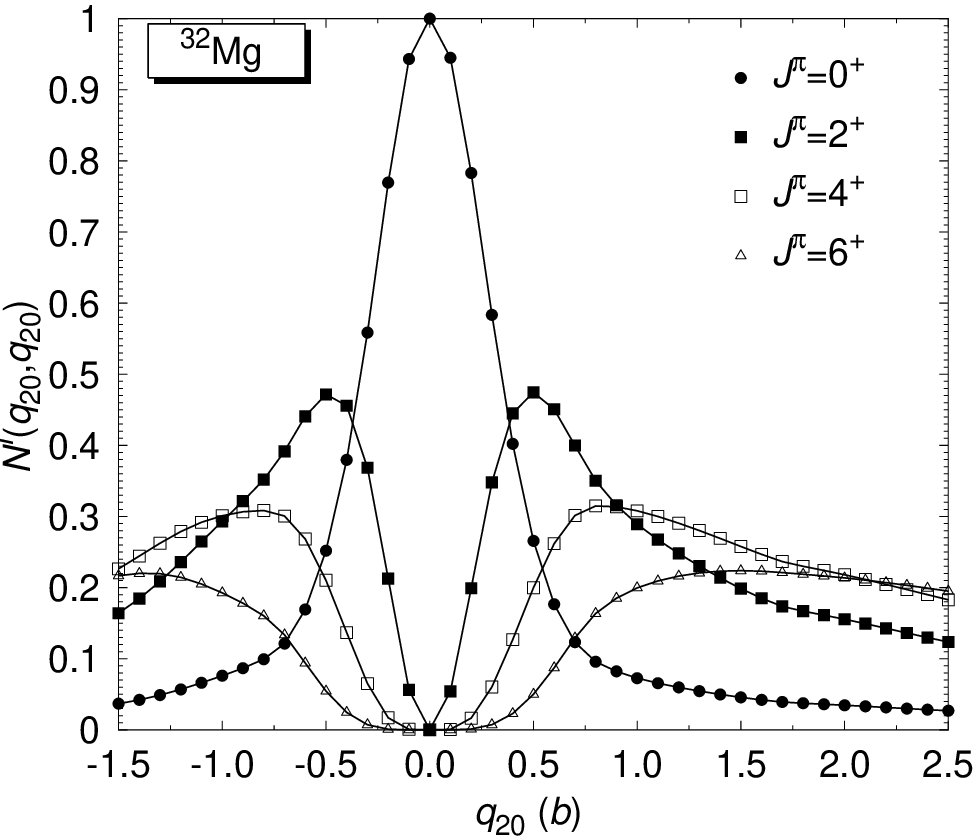}
\caption{Projected norm $\mathcal{N}^{I}(q_{20},q_{20})$ as a function of 
$q_{20}$ for the nucleus \nuc{32}{Mg}. For more details see the main text.}
\label{Fig_8_1_art}
\end{center} 
\end{figure}

The first noticeable fact that one can see is that, with the exception of the 
$I^{\pi}=0^{+}$ curves, several points around the spherical configuration  have
been omitted in the  AMPPES for the $I \ne 0$ curves. The reason for such
omission can be understood by looking at  Fig. \ref{Fig_8_1_art} where  the
behavior of the projected norm  $\mathcal{N}^{I}(q_{20},q_{20})=  \langle \phi
(q_{20})\mid \hat{P}_{00}^{I} \mid \phi(q_{20}) \rangle$ as a function of
$q_{20}$ for the nucleus \nuc{32}{Mg} is shown. From this figure  it becomes
clear that the omitted points, in this and the other  nuclei, correspond to
intrinsic configurations  with a very small value of the projected norm 
$\mathcal{N}^{I}(q_{20},q_{20})$. In those situations the projected energy is
the quotient of two very small quantities and therefore its evaluation is
affected by numerical inaccuracies that lead to erratic  values (deviations
from the smooth trend can be as large as a few MeV). Fortunately, due to the
smallness of the projected norm, these points can be  safely omitted since they
do not play any role in the configuration mixing calculations
\cite{ref_gg,ref_ii,ref_32S,ANGEL_24Mg,ref_ee,N=28} to be discussed later on. 
We also observe that no energy gain is obtained for the spherical 
configurations  and $I^{\pi}=0^{+}$. From a physical point of view this is the
expected behavior since these spherical  configurations are already pure
$0^{+}$ states as can  also be seeing in Fig. \ref{Fig_8_1_art} where we have 
$\mathcal{N}^{I=0}(0,0)=1$.

Comparing the MFPES and the  $I^{\pi}=0^{+}$ AMPPES of the  nuclei
\nuc{20}{Mg}, \nuc{28}{Mg}, \nuc{30}{Mg} and \nuc{32}{Mg} one can see   that
while at the mean field level spherical ground states are  obtained, once we
carry out angular momentum projection  two minima, one prolate and the other
oblate, appear. These minima  are very close in energy and the difference 
amounts to around 40, 80, and 49  keV in \nuc{20}{Mg}, \nuc{28}{Mg},
\nuc{30}{Mg}, respectively, and 637 keV in \nuc{32}{Mg}. These minima are
separated by a spherical barrier which is around  1, 3, and 2.6 MeV in 
\nuc{20}{Mg}, \nuc{28}{Mg}, \nuc{30}{Mg}, respectively, and 1.6 MeV in
\nuc{32}{Mg}.

It is noticeable to observe how the ground state of \nuc{32}{Mg} becomes
deformed after the inclusion of the rotational energy correction. The intrinsic
configuration corresponding to the shoulder seen in the MFPES at $q_{20}=0.8b$
has a big correlation energy coming from the restoration of the angular
momentum that is big enough as to overcome the energy difference with the
spherical configuration. The deformed intrinsic configuration at $q_{20}=0.8b$ 
corresponds to (see Fig.\ref{Fig_spe_32Mg}) a configuration in which two
neutrons from the $f_{7/2}$ shell have crossed the Fermi surface. In the shell
model language this is just a configuration where a pair of neutrons have been
promoted from the $sd$  to the $pf$ shell and this is the physical mechanism
invoked \cite{ref_kk,ref_mm} by the shell model practitioners to explain
deformation in this nucleus.

 All the other Magnesium isotopes, with the exception made of 
\nuc{26}{Mg} are prolate deformed at $I^{\pi}=0^{+}$. In fact, for 
increasing spin values the prolate minima become deeper than the 
oblate ones or the oblate minima are washed out. 

The nucleus \nuc{26}{Mg}, with its oblate intrinsic state at $I^{\pi}=0^{+}$ 
is the exception in all the Mg isotopes studied. As it was mentioned before,
the responsible for such oblate minimum is the  full occupancy of  the neutron
$d_{5/2}$ orbital which favors oblate deformations. The intrinsic state  of the
lowest lying $I^{\pi}=2^{+}$ state remains oblate deformed but already at
$I^{\pi}=4^{+}$ the underlying intrinsic state becomes prolate deformed.  The
spectroscopic quadrupole moment of the lowest $2^+$ state in \nuc{26}{Mg} is
known experimentally \cite{Endt98} to be -13.5 (20) fm$^2$ indicating that 
this is a prolate deformed state. However, the experimental low excitation
energy of the $0^+_2$ state in this nucleus (3.588 MeV, versus the 6.432 MeV for
the same quantity in  \nuc{24}{Mg}) is an indication of a strong shape
coexistence between the prolate and oblate solutions.

From the above discussed results we realize that the AMPPES for some nuclei and
some spin values show the phenomenon of  shape coexistence and therefore, a
configuration mixing calculation  is needed to disentangle the structure of
those states. Even in those situations where the AMPPES  show a relatively well
pronounced minimum it is always interesting to check its stability in the
framework of  the configuration  mixing calculation. The reason is that not
only the AMPPES but also  the collective inertia has to be considered in the
framework of  a dynamical treatment in order to determine the stability of a
given solution.  With this fact in mind, we have carried out  Angular Momentum
Projected Generator  Coordinate Method (AMPGCM) calculations along the lines 
described in subsection \ref{GCMSECTION} using the intrinsic  axial quadrupole
moment $q_{20}$ as generating coordinate. These  configuration mixing
calculations have been performed with  a mesh $\Delta q_{20}=10 fm^{2}$ which
was tested to be  accurate enough to describe, at least, the low-lying
spectrum  we are interested in this study.

\begin{figure}
\begin{center}
\includegraphics[angle=0,width=9cm]{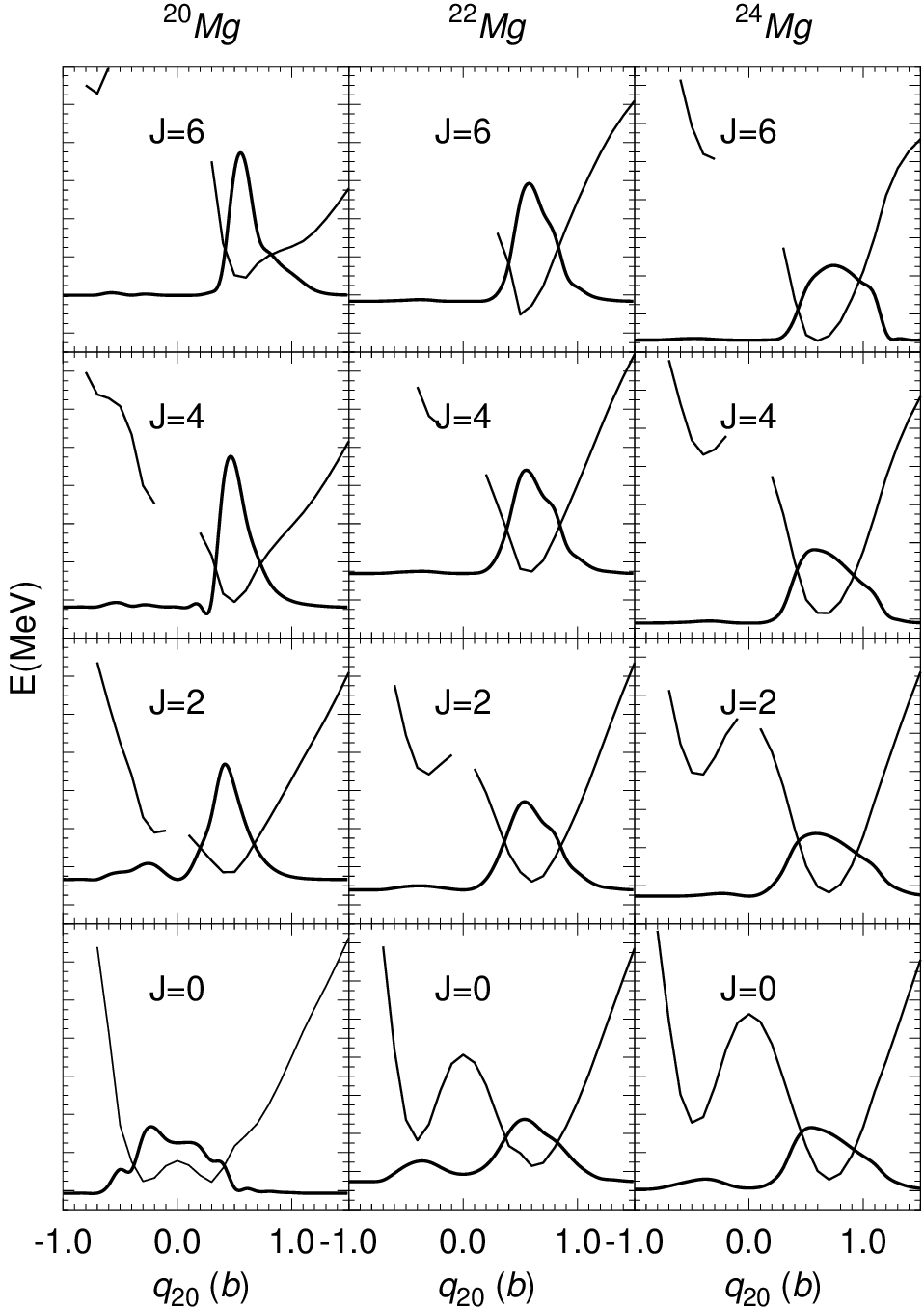}
\caption{Collective wave functions squared for  the ground states ($\sigma=1$)
and the  spin values $I^{\pi}=0^{+},2^{+},4^{+},6^{+}$ of the nuclei
\nuc{20,22,24}{Mg}. The corresponding  projected energy curve is also plotted
for each  spin value. The $y-axis$ scales  are in energy units and always span
an energy interval of $15$ MeV (minor ticks are $0.5$ MeV apart). The
collective wave functions  have also been plotted against the energy scale
after proper scaling  and shifting, that is, the quantity  $E^{I,\sigma} + 15
\times {\mid G^{I,\sigma=1}(q_{20}) \mid}^{2}$ is the  one actually plotted.
With this choice of scales we can read from the figure the energy gain due to
the quadrupole  fluctuations by considering the position of the wave functions'
tail relative to the projected curve.}
\label{Fig_9_art}
\end{center} 
\end{figure}

\begin{figure}
\begin{center}
\includegraphics[angle=0,width=9cm]{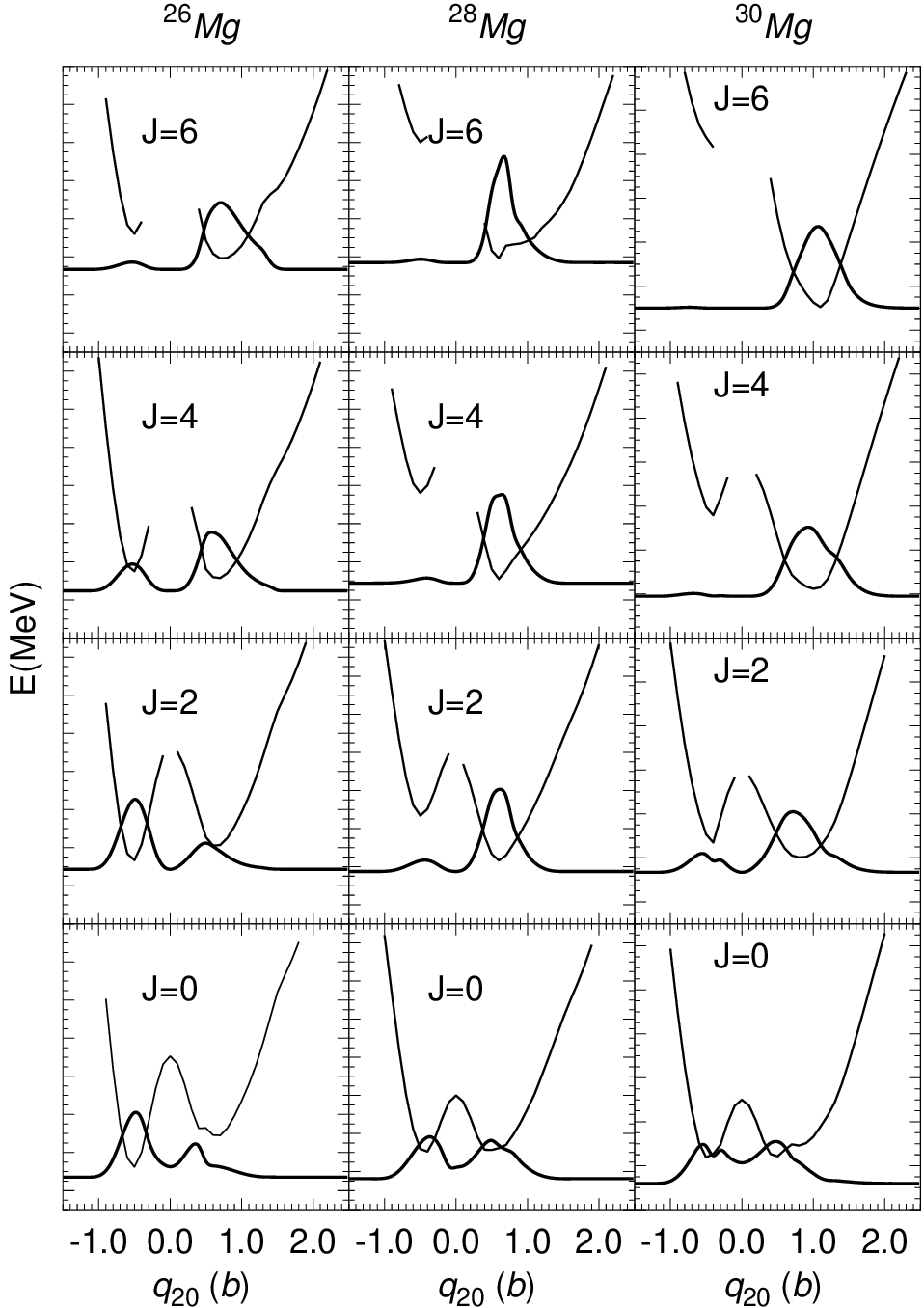}

\caption{The same as Fig. \ref{Fig_9_art} but for the nuclei \nuc{26,28,30}{Mg}.
The y axis scales span in this case an energy interval of$13$ MeV.}

\label{Fig_10_art}
\end{center} 
\end{figure}

\begin{figure}
\begin{center}
\includegraphics[angle=0,width=12cm]{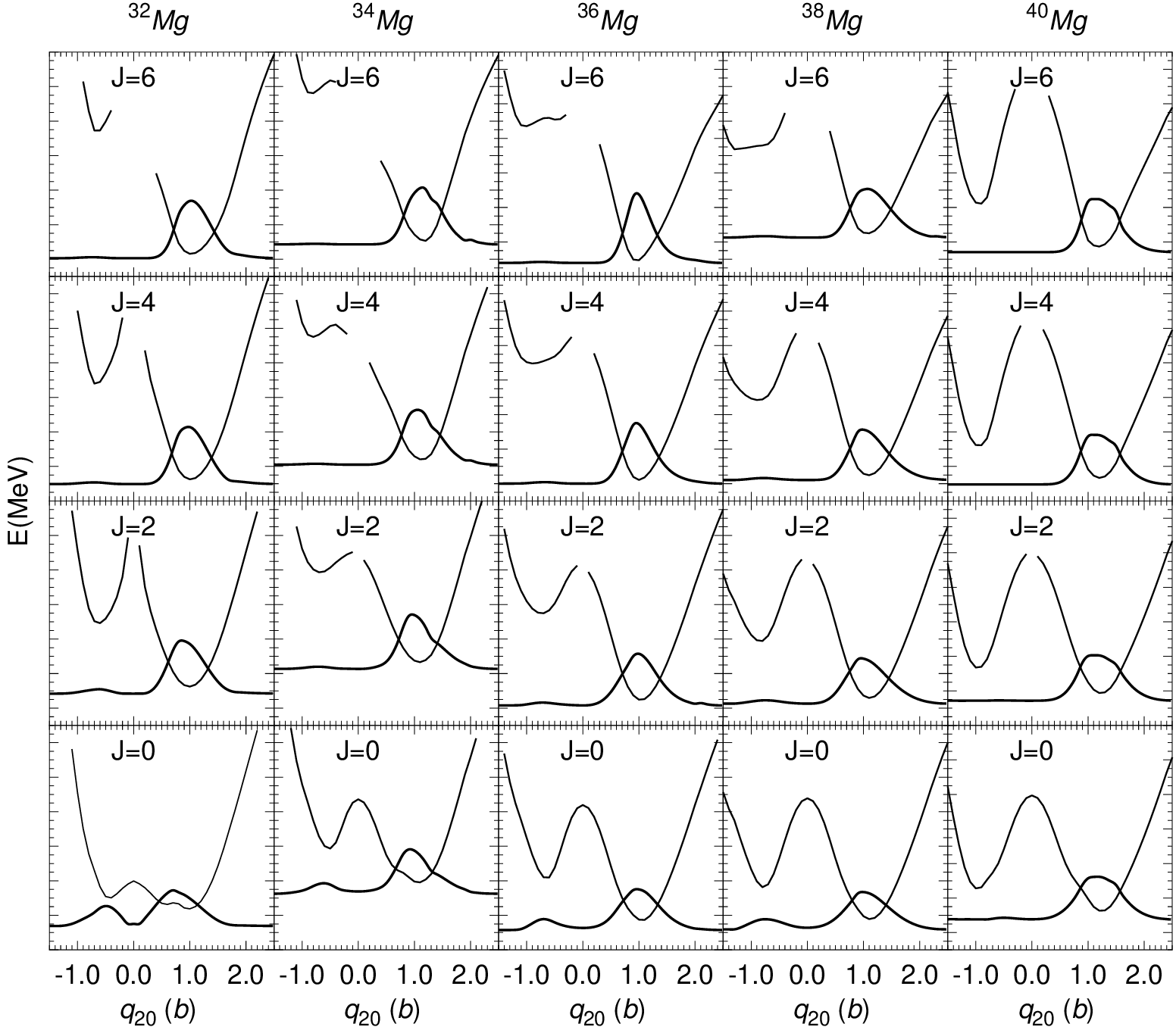}
\caption{The same as Fig. \ref{Fig_10_art} but for the nuclei \nuc{32-40}{Mg}.}
\label{Fig_11_art}
\end{center} 
\end{figure}

In Figs. \ref{Fig_9_art}, \ref{Fig_10_art}, and \ref{Fig_11_art} we  show the
ground  state ($\sigma=1$) collective wave functions squared  ${\mid
G^{I,\sigma=1}(q_{20}) \mid}^{2}$ for all the Magnesium  isotopes studied in
this paper up to $I^{\pi}=6^{+}$. We also plotted in each panel  the AMPPES for
the corresponding spin value. It is worth pointing  out that from the position
of the tails of the wave functions relative  to the projected energies (see
figure caption) we can read the energy  gain due to the quadrupole
fluctuations. In order to understand in a more quantitative way these 
collective wave functions  it is  convenient to analyze
\cite{ref_gg,ref_ii,ref_32S,N=28} the averages 
\begin{equation} 
{\bar{q}_{20}}^{I,\sigma}=  \int dq_{20} {\mid G^{I,\sigma}(q_{20}) \mid}^{2}
q_{20} \end{equation} 
that give a measure of the deformation of the underlying intrinsic states.

The $0_{1}^{+}$ wave function for \nuc{20}{Mg} shows a great admixture  between
the prolate and the oblate minima found in the  AMPPES. In fact, prolate and
oblate configurations have practically the same weight and therefore the ground
state of this nucleus is  spherical on the average. The  energy gain due to
quadrupole fluctuations is only 582 keV. As a result, the deformation effects
previously found in  the AMPPES are not stable once quadrupole fluctuations 
are taken into account and N=8 remains, on the average, as a  spherical magic 
number. On the other hand for spin values $I^{\pi} \ge 2^{+}$ the  ground state
collective wave functions are almost  inside the  prolate wells, with the
average deformations  $\bar{q}_{20}= 0.37 b$, $0.53 b$ and $0.64 b$ for the
states  $2_{1}^{+}$,$4_{1}^{+}$, and $6_{1}^{+}$ respectively.

For \nuc{22}{Mg} and  \nuc{24}{Mg} the ground state collective wave functions 
are well inside the prolate wells. This clearly shows on the one hand,  the
stability of the deformation effects found in the  corresponding AMPPES and on
the other that both systems are dominated by prolate deformations in the
considered spin range. The  average deformations are  $0.41 b$, $0.58 b$, $0.62
b$ and $0.63 b$ for the states  $0_{1}^{+}$, $2_{1}^{+}$, $4_{1}^{+}$, and
$6_{1}^{+}$ in  \nuc{22}{Mg}, while the corresponding values in \nuc{24}{Mg}
are $0.59 b$, $0.69 b$,  $0.70 b$, and $0.76 b$.

In the nucleus \nuc{26}{Mg}, both the $0_{1}^{+}$ and the  $2_{1}^{+}$ states
are slightly oblate deformed with  ${\bar{q}_{20}}^{I=0,\sigma=1}= -0.17 b$ and 
${\bar{q}_{20}}^{I=2,\sigma=1}= -0.16 b$ while for the state $4_{1}^{+}$ the
collective  wave function becomes  prolate deformed (i.e. a band crossing takes
place) with  ${\bar{q}_{20}}^{I=4,\sigma=1}= 0.39 b$ and 
${\bar{q}_{20}}^{I=6,\sigma=1}= 0.72 b$. In  both \nuc{28}{Mg} and \nuc{30}{Mg}
the ground state shows considerable  mixing between the oblate and prolate 
configurations. Here for the spin values  $I^{\pi} \ge 2^{+}$, the collective
wave functions are almost inside the  prolate wells. The average deformations
are  $0.50 b$, $0.60 b$ and  $0.67 b$ for \nuc{28}{Mg} for the states 
$2_{1}^{+}$, $4_{1}^{+}$, and $6_{1}^{+}$ while for \nuc{30}{Mg} the 
corresponding deformations for the same spin values are  $0.58 b$, $0.96 b$ 
and $1.09 b$. 

The oblate character of the $0^+$ and $2^+$ states in \nuc{26}{Mg}
already obtained without configuration mixing is preserved when it is included
(although in the latter case almost spherical configurations are obtained),
in contradiction with the experimental result (prolate character) extracted from the value of the
spectroscopic quadrupole moment of the $2^+$ state (-13.5 (20) fm$^2$). However,
our results predict a strong shape coexistence for those states in \nuc{26}{Mg}
as well as for the $0^+$ states of \nuc{28-30}{Mg}. A characteristic fingerprint
of shape coexistence comes from the position of the $0^+_2$ excited state: it is
expected to lie at a rather low excitation energy in those situations.
Experimentally \cite{Endt90,Endt98}, the excitation energy of the $0^+_2$ state is known in
\nuc{24}{Mg} (6.432 MeV), in \nuc{26}{Mg} (3.588 MeV) and in \nuc{28}{Mg}
(3.862 MeV). The sudden drop in the $0^+_2$ excitation energy in going from
\nuc{24}{Mg} to \nuc{26}{Mg} is a clear indication of shape coexistence in the
latter (and also in \nuc{28}{Mg}) nucleus. In our calculations, apart from the
shape of the collective wave functions, we get for those $0^+_2$ excitation
energies the values 5.675, 2.592 and 3.700 MeV for \nuc{24-28}{Mg},
respectively, that are a clear manifestation of shape coexistence in the last
two isotopes. The inclusion of dynamical pairing correlations may improve
our description of \nuc{26}{Mg}. Dynamical pairing correlations for neutrons 
will not change
much the oblate side due to the large single particle gap between the $d_{5/2}$
and the $s_{1/2}$ orbitals (see Fig. \ref{Fig_spe_32Mg}). However, in the
prolate side they can bring into the wave function the $f_{7/2}$ orbital 
which has a big
correlation energy coming from the restoration of the rotational symmetry. As a
consequence, the prolate side will gain more energy than the oblate one turning
the ground state of \nuc{26}{Mg} from slightly oblate to prolate. Although the
inclusion of dynamical pairing in the present calculations is rather cumbersome,
work along this line is in progress.

Now in \nuc{32}{Mg}, the $0_{1}^{+}$ collective wave function shows  a
significant mixing of the oblate and prolate configurations and  as a
consequence the deformation in the ground state is reduced from the $1.0b$
obtained taking the  absolute minimum in the $I^{\pi}=0^{+}$ AMPPES to
${\bar{q}_{20}}^{I=0,\sigma=1}= 0.43 b$ once  quadrupole fluctuations are taken
into account  via configuration mixing. This shows that the presence of a
deformed  ground state in this nucleus is the result of a subtle balance 
between the zero point corrections associated with the restoration  of the
rotational symmetry and the fluctuations in the collective  parameters (in our
case the axially symmetric quadrupole moment). On the  other hand, the presence
of a deformed ground  state indicates that for this nucleus N=20 is no longer a
magic number. The $2_{1}^{+}$, $4_{1}^{+}$, and $6_{1}^{+}$ wave functions in
this nucleus are inside the  prolate wells and the average deformations are
$0.88 b$, $1.01 b$ and $1.08 b$.

\begin{table}
\begin{center}
\label{TABLE_ESPECT}
\begin{tabular}{|c|c|c|c|}\hline
Nucleus & \multicolumn{3}{c|}{$Q^{spec}(I,\sigma=1)$} \\ \hline
  & I=2 & I=4 & I=6 \\ \hline \hline
\nuc{20}{Mg} & -12.59 & -23.04 & -30.60 \\ \hline
\nuc{22}{Mg} & -17.93 & -24.20 & -27.41  \\ \hline
\nuc{24}{Mg} & -19.69 & -25.14 & -28.80  \\ \hline
\nuc{26}{Mg} &   2.85 [-11.73] & -15.48 & -26.13  \\ \hline 
\nuc{28}{Mg} & -15.03 [-15.67] & -21.78 & -25.49 \\ \hline
\nuc{30}{Mg} & -13.19 [-12.4]  & -27.01 & -32.43  \\ \hline
\nuc{32}{Mg} & -19.15 [-18.1]  & -26.31 & -30.09   \\ \hline   
\nuc{34}{Mg} & -20.78 [-22.7]  & -27.59 & -31.27  \\ \hline
\nuc{36}{Mg} & -19.09 [-19.29] & -24.73 & -27.20 \\ \hline
\nuc{38}{Mg} & -18.59 [-19.45] & -24.48 & -27.22 \\ \hline
\nuc{40}{Mg} & -20.74 [-21.45] & -26.64 & -31.01  \\ \hline
\end{tabular}
\end{center}
\caption{Ground band spectroscopic quadrupole moments $Q^{spec}(I,\sigma=1)$ 
in $efm^{2}$ for 
$I^{\pi}=2^{+},4^{+},6^{+}$ in the nuclei \nuc{20-40}{Mg}. The Shell Model 
predictions from \cite{ref_nn,ref_pp} are shown in brackets.}
\end{table}

From \nuc{34}{Mg} to \nuc{40}{Mg}, the ground state collective wave functions 
become strongly prolate. The dynamical deformations of the $0_{1}^{+}$  states
are $0.79 b$, $0.77 b$, $0.79 b$, and  $1.17 b$ respectively. In all these
nuclei, the $\sigma =1$ collective  wave functions are well inside the prolate
wells in the whole spin range  considered. The dynamical deformation values for
the states $2_{1}^{+}$, $4_{1}^{+}$ and $6_{1}^{+}$ are  $1.0 b$, $1.13 b$,
$1.19 b$ in \nuc{34}{Mg}, $0.99 b$, $1.03 b$, $1.05 b$  in  \nuc{36}{Mg}, $1.03
b$, $1.10 b$, $1.14 b$ in  \nuc{38}{Mg} and  $1.23 b$, $1.25 b$, and $1.26 b$
in \nuc{40}{Mg}. The results show the  stability of the deformation effects in
Magnesium isotopes as we  move towards the dripline. Moreover, the presence of
a deformed ground state in \nuc{40}{Mg} also points towards the erosion of the
N=28 shell closure. Contrary to the N=20 case, the erosion of the N=28  shell
closure already appears at the mean field level and therefore we can say that
it is "stronger" than for the N=20 case. To summarize, it can be concluded
that, while N=8 remains a spherical  magic number in the Magnesium isotopic
chain, both N=20 and N=28 spherical shell closures do not remain.  

Here we will make a few comments on the results obtained in the  framework of
the configuration mixing approach for the nucleus  \nuc{40}{Mg} when the basis
is increased from $N_{Shell}=10$ to $N_{Shell}=11$. The effect on the projected
energy landscapes can be seen in the corresponding panel of Fig.
\ref{Fig_8_art} where the $N_{Shell}=11$ curves are plotted as dotted lines.
Although there are differences (specially at large absolute values of $q_{20}$
) between the  $N_{Shell}=10$ and $N_{Shell}=11$ curves, these differences are
almost independent of the considered spin. Therefore, we do not expect big
changes in the excitation energies of the ground state rotational band as is
the case: these excitation energies are, for all spin values, around 10 keV
higher for $N_{Shell}=11$ than for $N_{Shell}=10$. As a consequence, the
transition gamma ray energies remain unaltered by the increase of the basis
size. On the other  hand, the excitation energies (with respect to the true
ground state) of the members of the  excited rotational band decrease on the
average by 50 keV and therefore, as in the previous case, the intraband gamma
ray energies remain the same. From this  results we can conclude that our
calculations are well converged in terms  of the basis size.

The previous results suggest that all the considered nuclei 
are dominated by prolate deformations. This is clear 
from the results we have obtained for the average deformation 
${\bar{q}_{20}}^{I,\sigma=1}$ and also from the negative values of the 
ground band spectroscopic quadrupole moments presented in Table \ref{TABLE_ESPECT}.
With the exception of \nuc{26}{Mg}, our results for the spectroscopic 
quadrupole moments show a very good agreement with the Shell Model 
predictions of Ref.\cite{ref_nn,ref_pp} (these predictions are shown 
in brackets) for the spectroscopic quadrupole 
moments of the $2_{1}^{+}$ states. Experimentally, the $2^+$ spectroscopic
quadrupole moment of \nuc{24}{Mg} is -16.6 (6) $e^2$ fm$^2$ \cite{Endt98} and the
one of \nuc{26}{Mg} is -13.5 (20) $e^2$fm$^2$ \cite{Endt90}. For \nuc{24}{Mg} we
obtain a reasonable agreement with experiment (a 15 per cent discrepancy) whereas
for \nuc{26}{Mg} the disagreement is strong. In the latter case, we have already
traced back the disagreement to the strong shape coexistence obtained in our
calculations and also to the effect of dynamical pairing correlations (not
included in the present work).

\begin{figure}
\begin{center}
\includegraphics[angle=0,width=14cm]{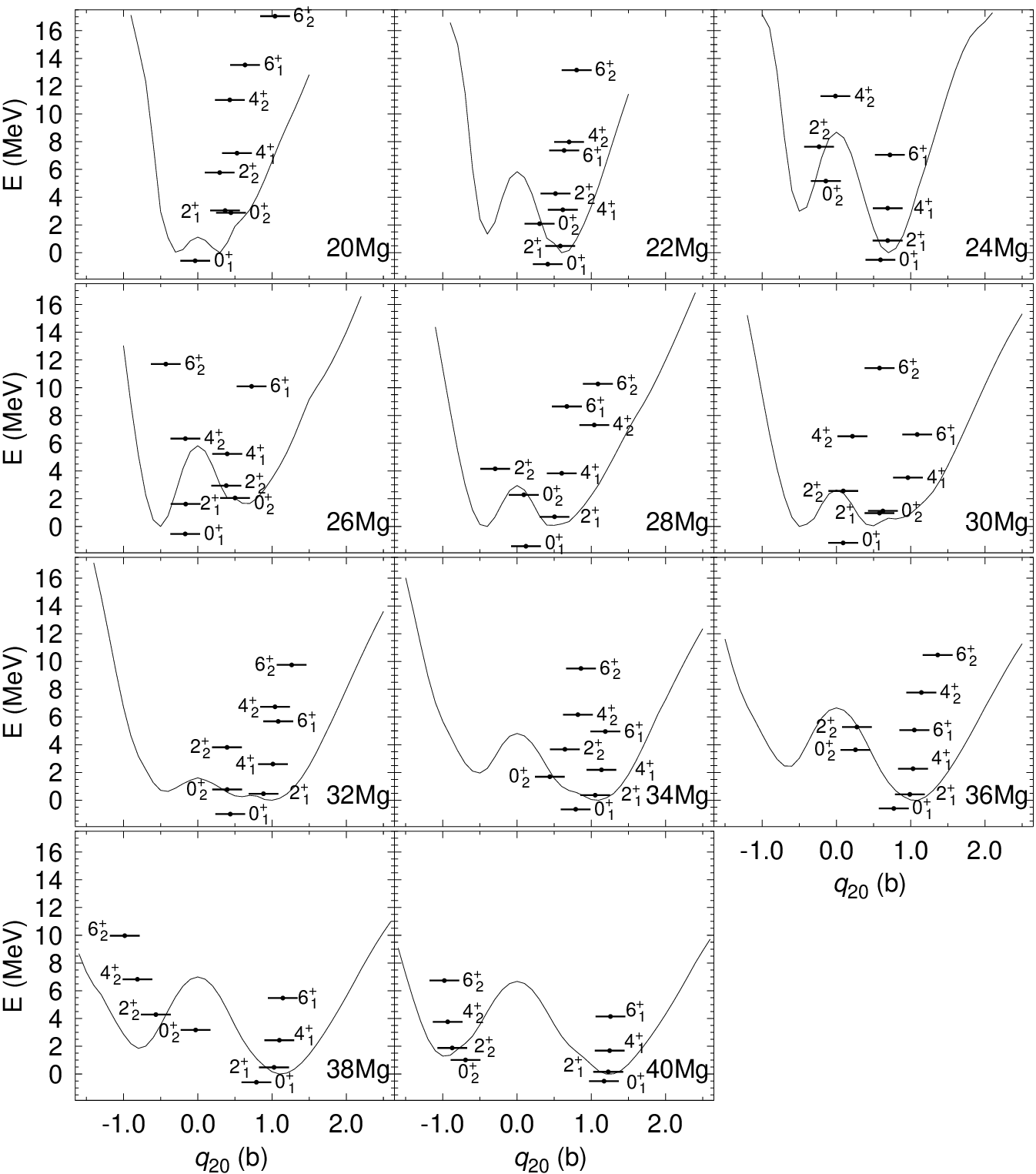}
\caption{The AMPGCM energies $E^{I,\sigma}$ for $\sigma=1$ and  $\sigma=2$ and
$I^{\pi}=0^{+},2^{+},4^{+},6^{+}$ are plotted in an energy versus quadrupole
moment diagram for the nuclei  \nuc{20-40}{Mg}. The quadrupole moment of each
AMPGCM state  is given by the average quadrupole moment 
${\bar{q}_{20}}^{I,\sigma}$. The AMPPES for  $I^{\pi}=0^+$ are also plotted to
guide the eye.}
\label{Fig_12_art}
\end{center} 
\end{figure}

The values obtained for the quadrupole moments ${\bar{q}_{20}}^{I,\sigma}$  of
the intrinsic states can be used to classify in terms of bands each of the
physical states  provided by the AMPGCM approach. In  Fig. \ref{Fig_12_art} we
have plotted the energies of the AMPGCM states $E^{I,\sigma}$ in a diagram of
energy versus quadrupole moment. Each energy  $E^{I,\sigma}$ is placed at a
$q_{20}$ value corresponding to  ${\bar{q}_{20}}^{I,\sigma}$. In addition, we
have plotted the  AMPPES for $I=0$ to guide the eye. Although, in the case of
the $\sigma =1$, many of the features observed in this  figure have already
been discussed, there we also show the results corresponding to the first
excited states ($\sigma=2$) provided by the  AMPGCM approach. Note that one of
the main advantages of such  representation is that the band structure of each
nucleus can be observed at a glance.

\begin{figure}
\begin{center}
\includegraphics[angle=0,width=8cm]{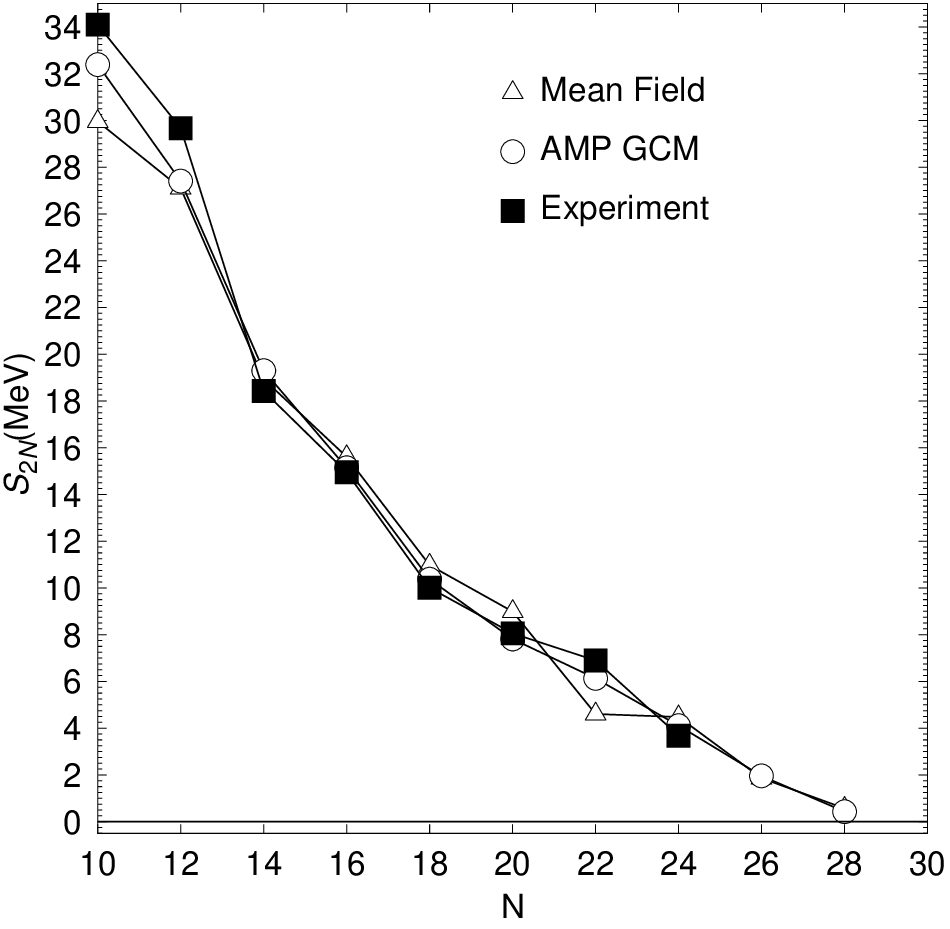}
\caption{The two neutron separation  energies $S_{2N}$ for Magnesium isotopes
as obtained in the framework of the AMPGCM are compared with the corresponding
mean field  values and also with the available experimental values  taken from
\cite{Audi.95}. The AMPGCM $S_{2N}$ are defined  as $E_{0_{1}^{+}}(N-2) -
E_{0_{1}^{+}}(N)$.}
\label{Fig_13_art}
\end{center} 
\end{figure}

In Fig. \ref{Fig_13_art} we compare the results for the AMPGCM two neutron 
separation energies $S_{2N}=E_{0_{1}^{+}}(N-2) - E_{0_{1}^{+}}(N)$ with the
corresponding mean field results (see subsection \ref{mean-field}) and also
with the available experimental values  \cite{Audi.95}. The AMPGCM binding
energy is the sum of the mean field binding energy of the intrinsic state plus
the energy gain due to the restoration of the rotational symmetry plus the
energy gain due to the configuration mixing. Therefore, the differences in the
two neutron separation energies obtained in the AMPGCM and the mean field are
due to the later two contributions. An analysis of those  contributions show
that the rotational energy correction is the main responsible for the
differences observed in the two neutron separation energies. The AMPGCM
$S_{2N}$ energies differ substantially from the mean field ones in the nuclei
\nuc{22}{Mg} and \nuc{34}{Mg} and are  much closer to the experiment.  On the
other hand, it is worth to remark that the nucleus \nuc{40}{Mg} remains  the
last bound isotope in the chain in both theoretical approaches.

\begin{figure}
\begin{center}
\includegraphics[angle=0,width=12cm]{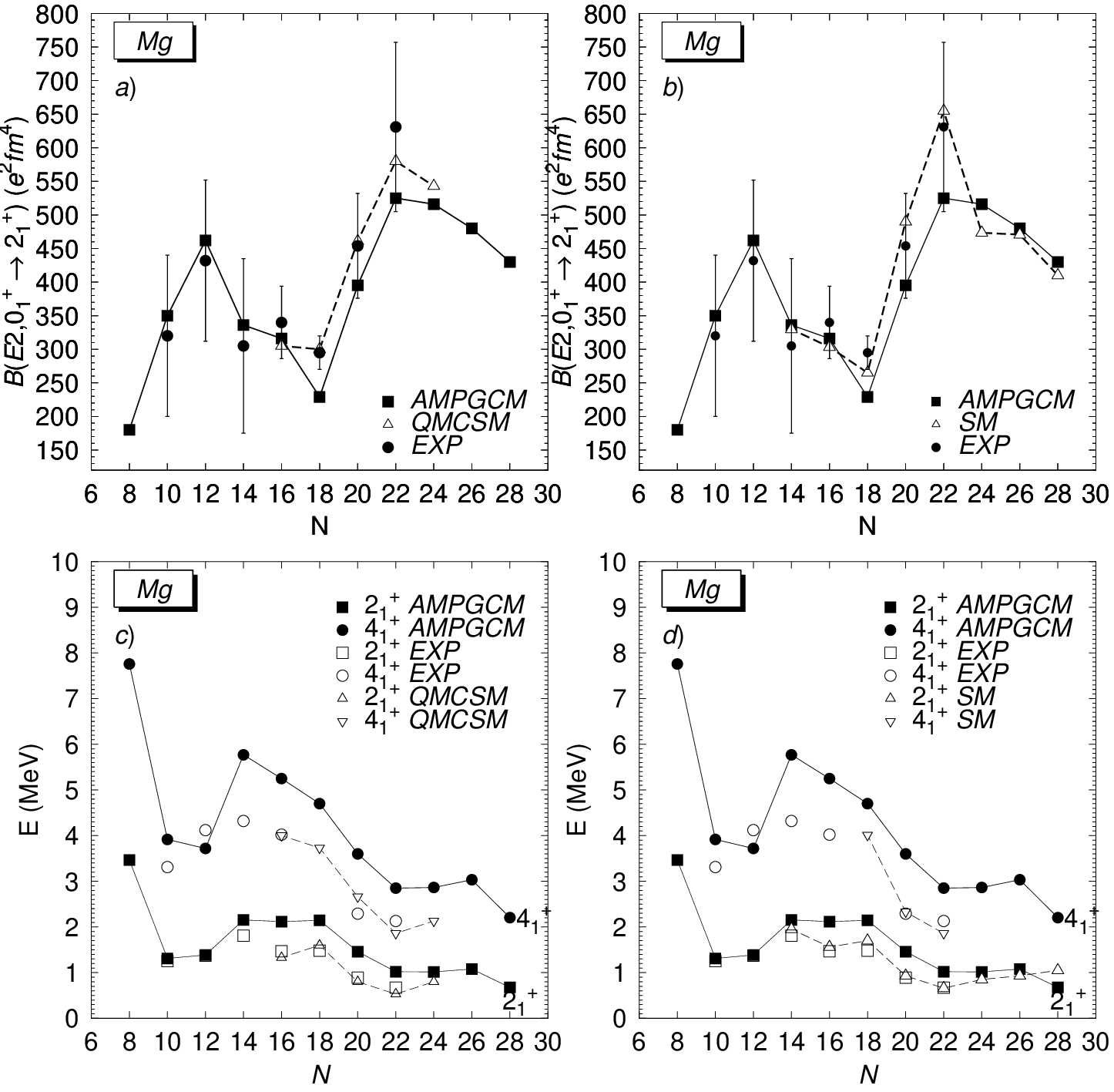}
\caption{The excitation energies of the states $2_{1}^{+}$ and  $4_{1}^{+}$
provided by the AMPGCM and the $B(E2,0^+_1\rightarrow 2^+_1)$ transition
probabilities in \nuc{20-40}{Mg} are compared with the available experimental
data \cite{Endt90,ref_p,ref_q,ref_t,ref_u,ref_v,BE2_26_28Mg} and with the
theoretical predictions of the Quantum Monte Carlo Shell Model \cite{ref_oo}
and the Shell Model \cite{ref_nn,ref_pp}.}
\label{Fig_14_art} 
\end{center} 
\end{figure} 

In Fig. \ref{Fig_14_art} the excitation energies  of the $2_{1}^{+}$ and 
$4_{1}^{+}$ states and the $B(E2,0^+_1\rightarrow 2^+_1)$ transition
probabilities  obtained in the framework of the AMPGCM are compared with the
available experimental values 
\cite{Endt90,ref_p,ref_q,ref_t,ref_u,ref_v,BE2_26_28Mg} and also with the 
predictions of the Quantum Monte Carlo Shell Model  \cite{ref_oo} and the Shell
Model  \cite{ref_nn,ref_pp}. Concerning the $B(E2)$ transition probabilities 
we clearly see, from panels a) and b), that  the agreement with the  available
experimental data is rather satisfactory and in most cases (with the exception
of \nuc{30}{Mg} where our prediction appears a little bit too low) our results
stay within the experimental error bars. On the other hand, our results  are
also consistent with the predictions of the  Quantum Monte Carlo Shell Model
\cite{ref_oo} and the  Shell Model \cite{ref_nn,ref_pp}. Our results, although
not as good as the SM or QMCSM ones, are very  satisfactory considering that
the parameters of the force have not been fitted to the region and/or the
physics of quadrupole collectivity and also that no effective charges have been
used in  our calculations  of the transition probabilities.

The calculated excitation energies for the $2_{1}^{+}$ and  $4_{1}^{+}$ states
are plotted in panels c) and d). They reproduce  quite well the experimental
isotopic trend and are again consistent with the theoretical trends predicted
by both the  Shell Model \cite{ref_nn,ref_pp} and the      Quantum Monte Carlo
Shell Model \cite{ref_oo}. In most cases, however, our values come too high as
compared with the experiment. In our  previous works \cite{ref_gg,N=28} we also
noticed  the same behavior in some N $\approx$ 20 and N $\approx$ 28 nuclei
(our predictions were too high as compared with the  experiment). Probably a
proper treatment of some missing  correlations will give the quenching factors
we need for a  much better agreement with the experiment. Although it is very 
difficult to assert before hand what are the missing  correlations, the mixing
of different $K$ values with a full triaxial angular momentum projection and a
beyond mean field  treatment of the dynamical pairing fluctuations  can be 
important ingredients for a more realistic description  of the nuclei studied
in this paper. Another source of discrepancy  could be related to the fact that
ours is a calculation of the  Projection After Variation (PAV) type instead of
the more  complete Projection Before the Variation (PBV). Having all  this in
mind and the free parameter character of our calculation we conclude that our
results  for the excitation energies of the  $2_{1}^{+}$ and  $4_{1}^{+}$
states show  a rather satisfactory agreement with the available experimental
and are consistent with other theoretical predictions. 

\section{Conclusions.}
 
With the aim to describe the phenomenology of quadrupole deformation in light
nuclei we have performed calculations with the Gogny force and in the framework
of  beyond mean field theories for the nuclei \nuc{20-40}{Mg}. First of all,
the results show the fundamental role played by the angular momentum projection
for a proper description of the physics under study. The effect of projection
in all the physical observables is so big that it can not be overlooked as it
has been common practice in many previous calculations. In addition, we also
find that the effect of configuration mixing is in most cases rather relevant.
From our results we conclude that for the Magnesium isotopes the N=8 shell
closure is preserved whereas for N=20 and N=28 deformed ground states appear in
the calculations. The three isotopes from \nuc{36Mg} to the drip line nucleus
\nuc{40}{Mg} are predicted to be prolate deformed in their ground states.
Concerning the excitation energies and $B(E2)$ transition
probabilities of the low-lying excited states we obtain a reasonable agreement
with experiment. The agreement is not as good as the one obtained with other
approaches like the Shell Model or the Quantum Monte Carlo Shell model. The
reason for that probably lies in the fact that our treatment of the problem,
although it contains the most important ingredients, is not as refined and
complete as the one of the SM and/or the QMCSM. In addition, the interaction
used has not been fitted to this specific region of the periodic table. The
later could be consider as a drawback of our calculations but we think it is
the other way around and in fact is a manifestation of the strong predicting
power of the Gogny interaction. It is rather satisfactory to obtain the results
presented in this paper with an interaction that is also able to reproduce, for
instance, the fission barrier heights of \nuc{240}{Pu}.

Finally, the consistency of the prescription used for the density dependent
term of the interaction in the present calculations beyond mean field has been
discussed in great detail.

\ack
This work has been supported in part by DGI, Ministerio de Ciencia y
Tecnolog\'\i a, Spain, under Project BFM2001-0184.    

\section{Appendix A: Calculation of transition probabilities.}

In this appendix we present  the basic formulas for the computation of angular
momentum projected  transition probabilities in the framework of the AMPGCM.
The starting point  is the transformation property of the multipole operators
$\hat{Q}_{\lambda \mu}$ under rotations

\begin{equation}
\hat{R}(\Omega) 
\hat{Q}_{\lambda \mu}
\hat{R}^{\dagger}(\Omega) =
\sum_{{\mu}^{'}} 
\mathcal{D}_{{\mu}^{'} \mu}^{\lambda}(\Omega) \hat{Q}_{\lambda {\mu}^{'}}
\end{equation}

Using the well known result for the product of two Wigner functions 
\cite{Varsh.88} as well as the definition  of Eq. (\ref{AMProj}) for the 
angular momentum projection operator and the property

\begin{equation} 
\hat{P}_{MK^{'}}^{I} \hat{P}_{K{'}M{'}}^{I'}= 
{\delta}_{I I{'}} {\delta}_{KK{'}} \hat{P}_{MM{'}}^{I}
\end{equation}
we obtain after some algebra the result

\begin{eqnarray}
\hat{P}_{K_{f}M_{f}}^{I_{f}} 
\hat{Q}_{\lambda \mu}
\hat{P}_{M_{i}K_{i}}^{I_{i}} & = & 
\langle I_{i}M_{i} \lambda  {\mu} \mid I_{f} M_{f} \rangle \nonumber \\
& \times & {\sum}_{\nu {\mu}^{'}} (-)^{{\mu}^{'} - \mu}
\langle I_{i} \nu  \lambda  {\mu}^{'} \mid I_{f} K_{f}\rangle
\hat{Q}_{\lambda {\mu}^{'}}
\hat{P}_{\nu K_{i}}^{I_{i}}.
\end{eqnarray}
With the definition of the projected wave functions of Eq. (\ref{PROJWF}) and the
previous result we obtain
\begin{equation}
\langle \Psi_{I_{f}M_{f}}(\vec{q}_{f}) \mid  
\hat{Q}_{\lambda \mu}
\mid \Psi_{I_{i}M_{i}}(\vec{q}_{i}) 
\rangle
=
\frac{
\langle I_{i}M_{i} \lambda {\mu} \mid I_{f} M_{f} \rangle
}{
\sqrt{2I_{f}+1}
}
\langle I_{f} \vec{q}_{f} \mid \mid \hat{Q}_{\lambda} \mid \mid
I_{i} \vec{q}_{i} \rangle
\end{equation}
with 

\begin{eqnarray} \label{RED_QLAMBDA}
\langle I_{f} \vec{q}_{f} \mid \mid & & \hat{Q}_{\lambda} \mid \mid
I_{i} \vec{q}_{i} \rangle
=
\frac{
(2I_{i}+1)(2I_{f}+1)
}
{8 {\pi}^{2}
}
(-)^{I_{i}- \lambda} 
\sum_{K_{i}K_{f} \nu {\mu}^{'}} (-)^{K_{f}} 
g_{K_{f}}^{I_{f}*}(\vec{q}_{f})
g_{K_{i}}^{I_{i}}(\vec{q}_{i})
\nonumber\\
& \times &
\left ( \begin{array} {ccc}
        I_{i} &    \lambda &  I_{f}  \\
        \nu   &  {\mu}^{'} & -K_{f}
        \end{array}  \right)
\int d \Omega \mathcal{D}_{\nu K_{i}}^{I_{i} *}(\Omega)
\langle \varphi(\vec{q}_{f}) \mid 
\hat{Q}_{\lambda {\mu}^{'}}
\hat{R}(\Omega)
\mid  \varphi(\vec{q}_{i})
\rangle	
\end{eqnarray}

Using now the expression (\ref{GCM_ANSATZ}) we get

\begin{eqnarray}
\langle \Phi_{I_{f}M_{f}}({\sigma}_{f}) \mid  
\hat{Q}_{\lambda \mu}
\mid \Phi_{I_{i}M_{i}}({\sigma}_{i}) 
\rangle =
\frac{
\langle I_{i}M_{i} \lambda {\mu} \mid I_{f} M_{f} \rangle
}{
\sqrt{2I_{f}+1}
}
\nonumber\\
\times
\int d\vec{q}_{i} d\vec{q}_{f}  f^{I_{f},{\sigma}_{f} *}(\vec{q}_{f}) 
\langle I_{f} \vec{q}_{f} \mid \mid \hat{Q}_{\lambda} \mid \mid
I_{i} \vec{q}_{i} \rangle
f^{I_{i},{\sigma}_{i} }(\vec{q}_{i}).
\end{eqnarray}
Finally, the expression for the
$B(E \lambda,{I_{i}} {{\sigma}_{i}} \rightarrow {I_{f}} {{\sigma}_{f}})$
transition probability is  written as 
\begin{eqnarray} \label{BE2_GCM_AM-cojo}
B(E \lambda & &,{I_{i}} {{\sigma}_{i}} \rightarrow {I_{f}} {{\sigma}_{f}})
=
\frac{e^{2}}{2I_{i}+1} \sum_{M_{i} M_{f} \mu}
\left |
\langle \Phi(I_{f},M_{f},{\sigma}_{f}) \mid  
\hat{Q}_{\lambda \mu}
\mid \Phi(I_{i},M_{i},{\sigma}_{i}) 
\rangle
\right | ^{2}
\nonumber\\
& = &
\frac{e^{2}}{2I_{i}+1}
\left |  
\int d\vec{q}_{i} d\vec{q}_{f}  f^{I_{f},{\sigma}_{f} *}(\vec{q}_{f}) 
\langle I_{f} \vec{q}_{f} \mid \mid \hat{Q}_{\lambda} \mid \mid
I_{i} \vec{q}_{i} \rangle
f^{I_{i},{\sigma}_{i} }(\vec{q}_{i})
\right | ^{2}
\end{eqnarray}

In the present work we are interested in the calculation  of transition
probabilities for axially symmetric  HFB  states labelled by the  quadrupole
deformation $q_{20}$. Taking advantage of the axial  symmetry of the intrinsic
wave function as well as  the selfconsistent symmetry $e^{-i \pi \hat{J}_{y}}$ 
we can simplify the above expressions as follows. First we have
\begin{equation}
\langle \varphi(q_{20,f}) \mid 
\hat{Q}_{\lambda {\mu}^{'}}
\hat{R}(\Omega)
\mid  \varphi(q_{20,i})
\rangle	
=
e^{i \alpha {\mu}^{'}}	
\langle \varphi(q_{20,f}) \mid
\hat{Q}_{\lambda {\mu}^{'}}
e^{-i \beta \hat{J}_{y}} 
\mid 
\varphi(q_{20,i}) \rangle	
\end{equation}
that leads to
\begin{eqnarray}
&\int & d \Omega \mathcal{D}_{QK_{i}}^{I_{i} *}(\Omega)
\langle \varphi(q_{20,f}) \mid 
\hat{Q}_{\lambda {\mu}^{'}}
\hat{R}(\Omega)
\mid  \varphi(q_{20,i})
\rangle	
= 4 {\pi}^{2} {\delta}_{Q -{\mu}^{'}} {\delta}_{K_{i}0}
\nonumber\\
& & \int_{0}^{\pi} d \beta \sin (\beta) d_{-{\mu}^{'}0}^{I_{i} *}(\beta)
\langle \varphi(q_{20,f}) \mid 
\hat{Q}_{\lambda {\mu}^{'}}
e^{-i \beta \hat{J}_{y}} 
\mid 
\varphi(q_{20,i}) \rangle
\end{eqnarray}
Applying this result to the expression of Eq. (\ref{RED_QLAMBDA}) we obtain
\begin{eqnarray} \label{RED_QLAMBDA_AXIAL}
& &\langle I_{f} q_{20,f}  \mid  \mid \hat{Q}_{\lambda} \mid \mid
I_{i} q_{20,i} \rangle
=
\frac{(2I_{i}+1)(2I_{f}+1)}{2}
(-)^{I_{i}- \lambda} 
\sum_{{\mu}^{'}}
\left ( \begin{array} {ccc}
        I_{i} & \lambda &  I_{f}  \\
        -{\mu}^{'} &  {\mu}^{'} & 0
        \end{array}  \right) 
\nonumber\\
& \times & \int_{0}^{\pi} d \beta \sin (\beta) d_{-{\mu}^{'}0}^{I_{i} *}(\beta)
\langle \varphi(q_{20,f}) \mid 
\hat{Q}_{\lambda {\mu}^{'}}
e^{-i \beta \hat{J}_{y}}
\mid  \varphi(q_{20,i})
\rangle	
\nonumber\\
& = &
(2I_{i}+1)(2I_{f}+1)
(-)^{I_{i}- \lambda} \frac{1+(-)^{I_{i}}}{2} 
\sum_{{\mu}^{'}}
\left ( \begin{array} {ccc}
        I_{i} & \lambda &  I_{f}  \\
        -{\mu}^{'} &  {\mu}^{'} & 0
        \end{array}  \right)
\nonumber\\
&\times &\int_{0}^{\frac{\pi}{2}} d \beta \sin (\beta) d_{-{\mu}^{'}0}^{I_{i} *}(\beta)
\langle \varphi(q_{20,f}) \mid 
\hat{Q}_{\lambda {\mu}^{'}}
e^{-i \beta \hat{J}_{y}}
\mid  \varphi(q_{20,i})
\rangle		 	
\end{eqnarray}
where we have, in the last line, reduced the integration interval to half the
original one.

Finally, from the previous expressions for the axially symmetric case, the  
$B(E2,{I_{i}} {{\sigma}_{i}} \rightarrow {I_{f}} {{\sigma}_{f}})$ in the
framework of the AMPGCM can be written as
\begin{eqnarray} \label{BE2_GCM_AM_axial}
B(E2 &,& {I_{i}} {{\sigma}_{i}} \rightarrow {I_{f}} {{\sigma}_{f}})
=
\frac{e^{2}}{2I_{i}+1} \nonumber \\
& \times & \left | 
\int dq_{20,i} dq_{20,f}  f^{I_{f},{\sigma}_{f} *}(q_{20,f}) 
\langle I_{f} q_{20,f} \mid \mid \hat{Q}_{2} \mid \mid
I_{i} q_{20,i} \rangle
f^{I_{i},{\sigma}_{i} }(q_{20,i})
\right |^{2}
\end{eqnarray}
with 

\begin{eqnarray} \label{RED_QLAMBDA_AXIAL_1}
\langle I_{f} q_{20,f} & & \mid \mid \hat{Q}_{2} \mid \mid
I_{i} q_{20,i} \rangle
=
(2I_{i}+1)(2I_{f}+1)
\sum_{{\mu}^{'}}
\left ( \begin{array} {ccc}
        I_{i} & 2 &  I_{f}  \\
        -{\mu}^{'} &  {\mu}^{'} & 0
        \end{array}  \right)
	\nonumber\\
&\times& 	
\int_{0}^{\frac{\pi}{2}} d \beta \sin (\beta) d_{-{\mu}^{'}0}^{I_{i} *}(\beta)
\langle \varphi(q_{20,f}) \mid 
\hat{Q}_{2 {\mu}^{'}}
e^{-i \beta \hat{J}_{y}}
\mid  \varphi(q_{20,i})
\rangle
\end{eqnarray}


\end{document}